\documentclass[usenatbib]{mnras}

\usepackage{newtxtext,newtxmath}

\usepackage[T1]{fontenc}
\usepackage{amsmath}

\DeclareRobustCommand{\VAN}[3]{#2}
\let\VANthebibliography\thebibliography
\def\thebibliography{\DeclareRobustCommand{\VAN}[3]{##3}\VANthebibliography}

\usepackage{ae,aecompl}
\usepackage{xparse,xspace}
\usepackage{cancel}
\usepackage{siunitx}
\usepackage{graphicx}	
\usepackage{amsmath}	
\usepackage{amsfonts}
\usepackage{graphicx}
\usepackage{color}
\usepackage{bm}
\usepackage{lipsum}
\usepackage{nccmath}
\usepackage{tabularx}
\usepackage{caption}
\usepackage{multicol}
\usepackage{graphicx}
\usepackage[export]{adjustbox}

\title[The Early Evolution of Magnetar Rotation I]{The Early Evolution of Magnetar Rotation I: Slowly Rotating ``Normal'' Magnetars}

\author[Prasanna et al.]{
Tejas Prasanna$^{1,2}$\thanks{E-mail: prasanna.9@osu.edu},
Matthew S. B. Coleman$^{3}$\thanks{E-mail: msbc@astro.princeton.edu},
Matthias J. Raives$^{2,4,5}$\thanks{E-mail: mraives@carnegiescience.edu},
\& Todd A. Thompson$^{2,4,1}$\thanks{E-mail: thompson.1847@osu.edu}
\\
$^{1}$Department of Physics, The Ohio State University, Columbus, Ohio 43210, USA\\
$^{2}$Center for Cosmology \& Astro-Particle Physics, The Ohio State University, Columbus, Ohio 43210, USA\\
$^{3}$Department of Astrophysical Sciences, Princeton, NJ 08540 USA\\
$^{4}$Department of Astronomy, The Ohio State University, Columbus, Ohio 43210, USA\\
$^{5}$The Observatories of the Carnegie Institution for Science, 813 Santa Barbara St., Pasadena, CA 91101, USA\\
}

\date{Accepted XXX. Received YYY; in original form ZZZ}

\pubyear{2022}

\begin{document}

\label{firstpage}
\pagerange{\pageref{firstpage}--\pageref{lastpage}}
\maketitle

\begin{abstract}
In the seconds following their formation in core-collapse supernovae, ``proto''-magnetars drive neutrino-heated magneto-centrifugal winds. Using a suite of two-dimensional axisymmetric MHD simulations, we show that relatively slowly rotating magnetars with initial spin periods of $P_{\star0}=50-500$\,ms spin down rapidly during the neutrino Kelvin-Helmholtz cooling epoch. These initial spin periods are representative of those inferred for normal Galactic pulsars, and much slower than those invoked for gamma-ray bursts and super-luminous supernovae. Since the flow is non-relativistic at early times, and because the Alfv\'en radius is much larger than the proto-magnetar radius, spindown is millions of times more efficient than the typically-used dipole formula. Quasi-periodic plasmoid ejections from the closed zone enhance spindown. For polar magnetic field strengths $B_0\gtrsim5\times10^{14}$\,G, the spindown timescale can be shorter than than the Kelvin-Helmholtz timescale. For $B_0\gtrsim10^{15}$\,G, it is of order seconds in early phases. We compute the spin evolution for cooling proto-magnetars as a function of $B_0$, $P_{\star0}$, and mass ($M$).  Proto-magnetars born with $B_0$ greater than $\simeq1.3\times10^{15}\,{\rm\,G}\,(P_{\star0}/{400\,\rm\,ms})^{-1.4}(M/1.4\,{\rm M}_\odot)^{2.2}$ spin down to periods $> 1$\,s in just the first few seconds of evolution, well before the end of the cooling epoch and the onset of classic dipole spindown. Spindown is more efficient for lower $M$ and for larger $P_{\star0}$. We discuss the implications for observed magnetars, including the discrepancy between their characteristic ages and supernova remnant ages. Finally, we speculate on the origin of 1E 161348-5055 in the remnant RCW 103, and the potential for other ultra-slowly rotating magnetars.
\end{abstract}




\section{Introduction}
\label{section:introduction}

In the seconds after the successful explosion of a massive star, a cooling proto-neutron star (PNS) radiates its gravitational binding energy in neutrinos \citep{Burrows1986}, heating the surface layers and driving a thermal wind \citep{Burrows1995, Janka1996}. Some neutron stars are born with large surface magnetic fields (``magnetars''; see \citealt{Kaspi2017}), with surface magnetic fields of order $10^{15}$\,G. Magnetar birth is common in the Galaxy, representing $\sim10-100$\% of neutron star births \citep{Beniamini2019}, and even modest progenitor rotation seems to be sufficient to produce magnetars during core-collapse \citep{White2022}. Estimates suggest that magnetar-strength magnetic fields could dominate the wind dynamics, affecting both the early spindown of magnetars and their nucleosynthesis  \citep{Thompson2003,Thompson2004,Bucciantini2006,Metzger2008,Thompson2018, Vlasov2014, Vlasov2017}. In combination with strong magnetic fields, rapid rotation may also be critically important for a subset of magnetar births, with spindown potentially producing normal supernovae \citep{Sukhbold2017}, super-luminous supernovae (SLSNe), and gamma-ray bursts (GRBs) \citep{Usov1992,Thompson1994,Wheeler2000,Zhang2001,Thompson2004,Bucciantini2006,Bucciantini2008,Bucciantini2009,Metzger2007,Komissarov2007,Kasen2010,Woosely2010,Metzger2011}.

Although much of the theoretical work on proto-magnetars and their evolution has been focused on GRBs and SLSNe, the prevalence of magnetars in the Galaxy and their otherwise normal supernova remnants \citep{Vink2008} suggest that the majority of magnetars are not born with the extreme rotation rates required to produce extreme explosions (initial periods of $\sim0.8-2$\,ms as in, e.g., \citealt{Thompson2004,Metzger2011,Margalit2018}). Prima facie, the finding that magnetars are common in the Galaxy with respect to massive star supernovae, and that their remnants do not exhibit signs of being anomalously energetic, suggests that very rapid rotation of order milliseconds is not required to produce their high magnetic fields (\citealt{Duncan1992,Thompson1993,Thompson1994,Raynaud2020}; see, e.g., \citealt{Barrere2022} for an alternate magnetar formation mechanism from fallback accretion).

Here, we explore the early angular momentum evolution of ``normal'' magnetars born with rotation rates representative of normal pulsars \citep{Faucher2006}, in the range $\sim50-400$\,ms. Following estimates by \cite{Thompson2004} and more detailed one-dimensional calculations by \cite{Metzger2007}, via a large set of two-dimensional magneto-hydrodynamic simulations, we find that these relatively slowly rotating magnetars spin down very rapidly during their first few seconds of existence, accompanying the early proto-neutron star cooling phase. These findings complement and have implications for the literature on the long-term magneto-thermal and spin evolution of the magnetar population of the Galaxy on kyr timescales \citep{Pons2011,Vigano2013,Mereghetti2015}.

In Section \ref{section:model}, we discuss our numerical simulations, including neutrino heating/cooling, the equation of state, boundary conditions, and our adopted computational reference frame. In Section \ref{section:results}, we present the results of our simulations, focusing on the spindown timescale of the PNS as a function of rotation period, polar magnetic field strength and neutrino luminosity. We also discuss the evolution of spin period of the PNS during the first $\sim6$\,s of evolution. In Section \ref{section:discussion}, we discuss the implications of rapid spindown of the PNS during the cooling phase. 

\section{Model}
\label{section:model}
We use the publicly available MHD code Athena++ \citep{Athena++} for our simulations, which we have configured to solve the following non-relativistic magneto-hydrodynamic (MHD) equations:
\begin{align}
          \frac{\partial \rho}{\partial t} + \nabla\cdot\left(\rho\mathbfit{v}\right)&=0\label{eq:continuity}\\
          \frac{\partial \left(\rho\mathbfit{v}\right)}{\partial t} + \nabla\cdot\left[\rho\mathbfit{vv}+\left(P+\dfrac{B^2}{2}\right)\mathbf{I}-\mathbfit{BB}\right]&=-\rho \frac{GM_{\star}}{r^2}\boldsymbol{\hat{r}}\label{eq:momentum}\\
          \frac{\partial E}{\partial t} + \nabla\cdot\left[\left(E+\left(P+\dfrac{B^2}{2}\right)\right)\mathbfit{v}-\mathbfit{B}\left(\mathbfit{B}\cdot\mathbfit{v}\right)\right]&=\dot{Q}\label{eq:energy}\\
          \frac{\partial \mathbfit{B}}{\partial t} -\nabla\times\left(\mathbfit{v}\times\mathbfit{B}\right)&=0,\label{eq:eulermag}
        \end{align}
where $M_{\star}$ is the mass of the PNS, $r$ is the radius from the center of the PNS, $\rho$ is the mass density of the fluid, $\mathbfit{v}$ is the fluid velocity, $E$ is the total energy density of the fluid, $P$ is the fluid pressure, $\dot{Q}$ is the heating/cooling rate (discussed in the following subsection), and $\mathbfit{B}$ is the magnetic field. Since the equations we solve are non-relativistic, our calculations are unreliable as the flow becomes relativistic. In general, we track the Alfv\'en speed and stop our calculation when it approaches the speed of light.

\subsection{Microphysics}
\label{micro}
The system of hydrodynamic equations is closed with an equation of state (EOS), and we have done so using the general EOS module in Athena++ \citep{Coleman2020}. Most of the characteristics of the wind are determined at temperatures $T\gtrsim 0.5$\,MeV and mass densities $\lesssim10^{12}$\,g cm$^{-3}$ . In these conditions, the outflow can be described by an equation of state containing non-relativistic baryons, relativistic electrons and positrons, and photons. We thus use the approximate analytic form of the general EOS from \cite{QW1996}. We have compared the results obtained using the approximate EOS with those obtained from the tabular Helmholtz EOS \citep{Timmes2000, Coleman2020}. We find that the results are approximately identical, both quantitatively and qualitatively. The run time of the simulations with the tabular Helmholtz EOS is significantly longer (at least $3-4$ times) than the simulations with the approximate EOS. Thus, the approximate analytic form of the general EOS is sufficient for our purposes. 

As discussed by \cite{QW1996}, the electron fraction $Y_{\rm e}$ varies across the wind profile. Assuming balance between the charged-current neutrino interactions at the neutrinosphere of the PNS, for the parameters in our calculations (see Section \ref{ICs}) which occur during the early cooling epoch, $Y_{\rm e}$ $\ll  0.1$ near the surface of the PNS which rapidly increases within a few PNS radius to reach an asymptotic value of approximately $0.45$, but varies as a function of time \citep{Vlasov2017}, depending on the electron and anti-electron neutrino luminosities and average energies \citep{QW1996}. For simplicity, we set $Y_{\rm e}=$\,constant in time and throughout the computational domain. To test this approximation, we have run 1D simulations with different fixed values of $Y_{\rm e}$ and compared with the results from the non-relativistic calculations of \cite{Thompson2001}. We find that using a constant $Y_{\rm e}$ does not affect the global wind properties such as the mass outflow rate ($\dot{M}$) and the adiabatic sonic radius (see Section \ref{diag}). However, the velocity profile and the neutrino heating/cooling rate (defined in the next paragraph) deviate from the results of \cite{Thompson2001} near the surface of the PNS. A future work will include a self-consistent calculation of $Y_{\rm e}$.

For neutrino heating and cooling, we consider the two most important interactions between the neutrinos and the wind material: charged-current neutrino absorption and electron capture on free nucleons: $\nu_e+n \rightleftharpoons p+e^{-}$ and $\bar{\nu}_e+p \rightleftharpoons n+e^{+}$. We use the optically-thin specific heating rate and cooling rate expressions from \cite{QW1996} as the $\dot{Q}$ function in equation \ref{eq:energy}. At a given radius $r$, neutrinos emitted from the neutrinosphere are visible only within the solid angle subtended by the neutrinosphere at $r$. We incorporate this effect by using the geometric factor as in \cite{QW1996}. We incorporate neutrino heating/cooling into the code through a source function which adds this energy to the wind energy. 

The approximations in our calculations are similar to those used in \cite{Thompson2018}. Bulk features of the wind are reproduced well with these approximations. For the purposes of this paper, we  neglect General Relativistic effects. GR effects have been considered in several previous works (e.g.,  \citealt{Cardall1997,Otsuki2000,Wanajo2001,Thompson2001}).  GR increases the effective gravitational potential, and the associated gravitational redshift terms decrease the heating rate, both leading to lower $\dot{M}$, larger entropy, and shorter expansion timescales \citep{Cardall1997}. The neutrino heating is augmented by geodesic bending in GR \citep{Salmonson1999}, but the decrease in the heating rate due to gravitational redshift dominates \citep{Cardall1997, Thompson2001}. We will consider more detailed microphysics and GR effects in a future work.

\subsection{Rotating reference frame}
To simplify the magnetic field boundary conditions at the inner boundary, we perform our simulations in a frame rotating with the PNS at an angular velocity $\boldsymbol{\Omega}_{\star}=\Omega_{\star} \boldsymbol{\hat{z}}$. Henceforth, primed coordinates refer to the quantities in the rotating frame and unprimed coordinates refer to the quantities in the lab frame. We have the following equation connecting the velocity in the lab and the rotating frame:
\begin{equation}
    \mathbfit{v}= \mathbfit{v}^{\prime}+ \  \boldsymbol{\Omega}_{\star} \times \mathbfit{r}.
\end{equation}
The magnetic field remains invariant under frame transformation in Galilean relativity.

We must include the pseudo forces (Coriolis and centrifugal forces here) since we are in a non-inertial reference frame. The centrifugal force also contributes to the the total energy of the outflow. We incorporate these into our simulations as an extra source term following Appendix A of \cite{Zhu2021}. In a subset of our calculations, we evolve the angular velocity of the PNS self-consistently with time. In these calculations (details of this in Section \ref{section:results}), we include the Euler force in the source term. 

\subsection{Initial conditions}
\label{ICs}
We start from the surface of the PNS with radius $R_{\star}=12$\,km. We use a spherical coordinate system with the outer boundary of the grid at a radius of $10^{3}-10^{4}$\,km. The base density of the PNS is set at $\rho_0 = 1.44 \times 10^{12}$\,g cm$^{-3}$. We tested our 1D simulations with different base radii ($10-12$\,km) and base densities $10^{11}-10^{13}$\,g cm$^{-3}$. We find that the mass outflow rate is not very sensitive to the base density over the $\rho_0$ range ($10^{11}-10^{13}$\,g cm$^{-3}$) in our calculations, the variation being less than 10\% with at least 1024 radial zones. All else fixed, the mass outflow rate roughly increases as $R_{\star}^{5/3}$ \citep{QW1996}.

We reference our simulations by the electron anti-neutrino luminosity $L_{\rm \bar{\nu}_e}$. $L_{\rm \bar{\nu}_e}$ starts at $\sim 10^{52}$\,ergs s$^{-1}$ at the time of the SN and decreases over the cooling timescale. For this paper, we consider $L_{\rm \bar{\nu}_e}$ ranging between $2\times 10^{52}$\,ergs s$^{-1}$ and $1.5\times 10^{5}$\,ergs s$^{-1}$. This range corresponds to the neutrino luminosity at various times after a supernova depending on the cooling model \citep{Burrows1986, Pons1999, Li2021}. In the \cite{Pons1999} cooling model, the luminosity range we consider corresponds to the first $\sim 6$\,s of PNS evolution. The electron neutrino luminosity is assumed to be given by $L_{\rm \nu_e}=L_{\rm \bar{\nu}_e}/1.3$ \citep{Thompson2001}. However, the ratio of the luminosities can be different  from this value and change as a function of time depending on the PNS properties and the cooling calculation \citep{Vlasov2017}. We do not consider neutrinos of other flavors for simplicity. We define the mean neutrino energy $\langle\epsilon_{\nu}\rangle$ and variance $\langle\epsilon_{\nu}^2\rangle$ in terms of the $n$th moment of neutrino energy distribution: $\langle\epsilon_{\nu}\rangle=\frac{\langle E^3\rangle}{\langle E\rangle}$ and $\langle\epsilon_{\nu}^2\rangle=\frac{\langle E^5\rangle}{\langle E^3\rangle}$. These quantities are related through the Fermi integrals: $\langle\epsilon_{\nu}^2\rangle=\langle\epsilon_{\nu}\rangle^2 \frac{F_5(0)}{F_3(0)}\left(\frac{F_2(0)}{F_3(0)}\right)^2$ \citep{Thompson2001}. Neutrino mean energy first increases and then decreases during the first few seconds ($\sim 3-5$\,s depending on the PNS mass) after the supernova \citep{Pons1999}. Unless otherwise stated, the results presented here have  $\langle\epsilon_{\rm \bar{\nu}_e}\rangle=14$\,MeV and $\langle\epsilon_{\rm \nu_e}\rangle=11$\,MeV and we assume that the mean energy remains constant as luminosity evolves. We initialize the simulations (both 1D and 2D, irrespective of the neutrino luminosity used in the simulation) using data from a spherically symmetric non-rotating 1D wind model corresponding to $L_{\rm \bar{\nu}_e}=10^{52}$\,ergs s$^{-1}$  \citep{Thompson2001}.  

PNSs with ``slow'' rotation ranging from $50$\,ms to $400$\,ms are the focus of this work. We will consider rapid rotation with periods $\sim 1$\,ms in a future work. The velocity in the $\phi$ direction is initialized using angular momentum conservation: 
\begin{equation}
    v_{\phi}^{\prime}(r)= r\Omega_{\star}\sin \theta\left(\frac{R_{\star}^2}{r^2}-1\right).
\end{equation}
Although this initialization is incorrect for simulations with magnetic field, we find that the system quickly relaxes to the correct state. 

In this work, we assume a spherical inner boundary and that the neutrino energy is independent of latitude. Centrifugal forces due to rapid rotation can deform the spherical emitting surface \citep{Petri2022}. Rapid rotation can also result in neutrino energy being a function latitude.   

For the 2D simulations including a  magnetic field, the initial conditions for the magnetic field are set assuming a dipole magnetic field with polar magnetic field $B_0$. The dipole field is specified using the following magnetic vector potential:
\begin{equation}\label{vector_pot}
    \mathbfit{A}\left(r,\theta,\phi \right)=\frac{B_0}{2}\frac{R_{\star}^3}{r^2}\left(\hat{\phi} \ {\rm sin} \ \theta \right).
\end{equation}

Provided that the initial conditions are sensible enough and the simulation is run long enough, the final state is not sensitive to the initial conditions. Most of the models (unless otherwise stated) assume a PNS with a mass of $1.4$ M$_{\odot}$. We discuss the effects of the PNS mass in Section \ref{section:results}.

\subsection{Resolution of the grid}
For the 1D simulations, we use a logarithmically spaced grid with $N_r$ number of radial zones. In the 1D case in Athena++, it is important to set the extent of $\theta$ symmetrically about $\frac{\pi}{2}$ to avoid the geometric source terms.  For the 2D simulations, we use a logarithmically spaced grid with $N_r$ zones in the radial direction and a uniformly spaced grid with $N_{\theta}$ zones in the $\theta$ direction ($0 \leq \theta \leq \pi$). 

The results are independent of the resolution as long as the grid is sufficiently well-resolved. For example, in our 1D fiducial models with $L_{\bar{\nu}_{\rm e}}=8\times 10^{51}$\,ergs s$^{-1}$, we find that $\dot{M}=3.61\times 10^{-4}$\,M$_{\odot}$ s$^{-1}$ with 1024 radial zones. The deviation from this value is 0.2\%, 1.6\% and 13\% at 512, 256, and 128 radial zones, respectively. Using too low a radial resolution gives systematically inaccurate mass outflow rate. Insufficient $\theta$ resolution in our 2D MHD simulations affects the dynamical plasmoid mass ejections (described in Section \ref{section:results}). For instance, dynamical eruptions are absent for $\theta$ resolution less than 128 zones at $B_0 \geq 4\times 10^{15}$\,G. We have run 1D simulations with $N_r$ = 128, 256, 512, 1024, and 2048. The 2D simulations have $(N_r,\,N_{\theta}) = (128,64), (256,128), (512,256)$ and $(1024,512)$. High luminosity models require comparatively fewer radial zones to achieve the required accuracy. For example, in our 1D non-rotating-non-magnetic (NRNM) simulations at $L_{\rm \bar{\nu}_e}=8\times 10^{51}$\, ergs s$^{-1}$, we require at least 256 radial zones to obtain $\dot{M}$ within 2\% of the corresponding value at 1024 radial zones, while at $L_{\rm \bar{\nu}_e}=1\times 10^{51}$\, ergs s$^{-1}$, we require at least 512 radial zones to achieve the same accuracy. This is because the decrease in density per radial zone near the surface is larger at lower neutrino luminosities (i.e., the density scale height is smaller; e.g., \citealt{Thompson2001}).    

\subsection{Boundary conditions}
\label{BCs}
As in earlier wind calculations \citep[e.g.][]{Thompson2001}, the inner boundary temperature $T_0$ is set by equating the neutrino heating and cooling rates. For $L_{\rm \bar{\nu}_e}=8\times 10^{51}$\,ergs s$^{-1}$ and $Y_{\rm e}=0.45$, $T_0 \sim 4$\,MeV.

At the inner boundary we set the following boundary condition for the density (derivation in Appendix \ref{app:BCs}): 
\begin{equation}\label{densBC}
\begin{split}
    \rho(r,\theta)=\rho_0\exp\left[\frac{GM_{\star}m_{\rm n}}{kT_0} \left(\frac{1}{r} -\frac{1}{R_{\star}}\right)\right] \exp\left[\frac{m_{\rm n}r^2\Omega_{\star}^2\sin^2\theta }{2kT_0}\right] \\
    \times \exp \left[\frac{-m_{\rm n} R_{\star}^2\Omega_{\star}^2}{2kT_0}\right],
    \end{split}
\end{equation}
where we have assumed that the inner boundary is at a constant temperature and that ideal nucleons dominate the pressure in the EOS (see Appendix \ref{app:BCs} and \citealt{QW1996}). In equation \ref{densBC}, $m_{\rm n}$ is the average mass of a nucleon and $\rho_0=1.44 \times 10^{12}$\,g cm$^{-3}$ is the surface density of the PNS in the non-rotating case. 
The velocity vector is set to zero at the inner boundary. Although $v_r=0$ is not consistent with a steady mass outflow, $v_r=0$ satisfies the axisymmetry condition (see Appendix \ref{app:BCs}), works well, and gives consistent results with other choices for the inner velocity boundary condition. Similar to our experience in other contexts, we find that the wind parameters are not sensitive to the details of the $v_r$ boundary condition. For example, letting the value of $v_r$ ``float'', by setting $v_r$ at the inner boundary equal to the value at the first active zone and setting $v_r$ using $\rho v_r r^2=$\,constant also give the same results. We note that the $\theta$ dependence of the density boundary condition in equation (\ref{densBC}) is crucial to get angular momentum conservation.

In order to keep the magnetic field with field strength $B_0$ at the poles (see Section \ref{ICs} and equation \ref{vector_pot}) constant in time at the surface of the PNS in the rotating frame, we set the electric field components $E_{\theta}$ and $E_{\phi}$ to zero at the inner boundary which follow from Maxwell's equations.

The boundary conditions at the outer boundary are set by conserving the angular momentum and the mass outflow rate:
\begin{align}
    \rho(r,\theta)&= \rho(r_{\rm max},\theta)\left(\frac{r_{\rm max}}{r}\right)^2 \\
    v_r(r,\theta)&=v_r(r_{\rm max},\theta) \\
    v_{\theta}(r,\theta)&=v_{\theta}(r_{\rm max},\theta)\frac{r_{\rm max}}{r} \\
    v_{\phi}(r,\theta)&=v_{\phi}(r_{\rm max},\theta)\frac{r_{\rm max}}{r} \ + \ \Omega_{\star}\sin \theta \left( \frac{r_{\rm max}^2-r^2}{r}\right),
\end{align}
where $r_{\rm max}$ is the radius of the last active zone on the grid. 

At both the inner and outer boundaries, we enforce constant magnetic fields by copying the magnetic field values from the nearest active zone into the ghost zones. Since we enforce a constant magnetic field with time at the surface of the PNS in the rotating frame, the magnetic field at the inner boundary remains constant in time as well, with the polar magnetic field strength being $B_0$.

\section{Results}
\label{section:results}
\subsection{Diagnostic quantities}
\label{diag}
For time-steady inner boundary conditions, we expect the mass outflow rate $\dot{M}$ to be a constant of the wind. From the continuity equation (eq. \ref{eq:continuity}), $\dot{M}$ is given by the following surface integral over a sphere:
\begin{equation}
\label{Mdot}
    \dot{M} \left(r \right)= \oint_S r^2 \rho v_r d\Omega.
\end{equation}
A principal output of our simulations is the angular momentum loss rate. The $z$-component of the angular momentum flux is given by the following integral over a closed spherical surface \citep{Vidotto2014}:
\begin{equation}
\label{Jdot}
    \dot{J}\left(r \right)= \oint_S \left[-\frac{B_rB_{\phi}r\sin \theta}{4\pi}+\rho v_r v_{\phi}r\sin \theta\right]r^2 d\Omega.
\end{equation}
The total angular momentum of the star is roughly $J=\frac{2}{5}MR_{\star}^2\Omega_{\star}$. We define the spindown time of the PNS as:
\begin{equation}
\label{tauj}
    \tau_{\rm J}=\frac{J}{\dot{J}}.
\end{equation}
The energy flux is given by the following surface integral (we generalize the definition in \cite{Metzger2007} to two and three dimensions):
\begin{align}
\label{Edot}
    \dot{E}\left(r \right)&= \oint_S r^2 \rho v_r\left[\frac{1}{2}\left(v_r^2+v_{\theta}^2+v_{\phi}^2\right) -\frac{rB_rB_{\phi}\Omega_{\star}\sin\theta}{\rho v_r}\right.  \\  \nonumber &\qquad -\left.\frac{GM_{\star}}{r} +e+\frac{P}{\rho}\right] d\Omega,
\end{align}
where $e$ is the specific internal energy of the outflow.

We measure $\dot{J}$ and $\dot{M}$ at a radius of $50$\,km. We note that the time-average values of $\dot{J}$ and $\dot{M}$ are independent of the radius at which they are measured as long as the measurement is not within $\sim 50$ zones of the outer boundary and not within $\sim 20$ zones of the inner boundary. We measure the asymptotic $\dot{E}$ at a radius of $1000$\,km. Asymptotic $\dot{E}$ can be measured at any radius not within $\sim 50$ zones of the outer boundary and large enough where the the neutrino heating rate is at least ten times smaller than its peak value \citep{Metzger2007}. The number of inner and outer zones to be excluded for the measurement of physical quantities is a function of radial resolution. The above exclusion of 50 outer zones and 20 inner zones is at 512 radial zones with the outer boundary at $3000$\,km. Rapid frame rotation near the outer boundary distorts the measurement of the physical quantities as a result of which we have to be careful not to measure the physical quantities too close to the outer boundary. The distortion near the outer boundary decreases with increasing number of radial zones. The measurement should also not be too close to the inner boundary where boundary conditions can produce resolution-dependent effects on the profiles. We note that all the physical quantities have to be averaged over time. The time average over integer number of plasmoids is necessary in order to account for the variations due to plasmoids (described in Section \ref{MHD results}). In the absence of plasmoids, at a given rotation period of the PNS, $\dot{J}$ and $\dot{M}$ are constant in radius and time.

We define three important surfaces related to the magnetosonic speeds. At the adiabatic sonic surface, the poloidal wind speed ($v_r^2+v_{\theta}^2$) is equal to the adiabatic sound speed ($c_{\rm s}$). At the Alfv\' en surface, the poloidal wind speed is equal to the Alfv\' en speed:
\begin{equation}
    v_r^2+v_{\theta}^2=v_{\rm A}^2=\frac{B_r^2+B_{\theta}^2}{4\pi \rho}.
\end{equation}
The fast/slow magnetosonic speed is given by:
\begin{equation}
    v_{\pm}^2=\frac{1}{2}\left(v_{\rm A}^2+c_{\rm s}^2\pm \sqrt{\left(v_{\rm A}^2+c_{\rm s}^2\right)^2-4v_{\rm A}^2c_{\rm s}^2 {\rm cos}^2 \Theta}\right),
\end{equation}
where $\Theta$ is the angle between the magnetic field and the direction of wave propagation. At the fast/slow magnetosonic surface, the poloidal speed is equal to the fast/slow magnetosonic speed. For the purpose of determining the fast magnetosonic surface, we assume that the magnetosonic waves propagate radially. Thus, we have ${\rm \cos} \ \Theta=\mathbfit{B}\cdot \hat{\mathbfit{r}}/B$, where $B$ is the magnitude of the magnetic field. 

Tables \ref{table1}, \ref{table2}, \ref{table3} and \ref{table4} give the values of important physical quantities including spindown timescale $\tau_{\rm J}$ (eq. \ref{tauj}), angular momentum flux $\dot{J}$ (eq. \ref{Jdot}), mass flux $\dot{M}$ (eq. \ref{Mdot}), average Alfv\'en radius $\langle R_{\rm A}\rangle$, average adiabatic sonic radius $\langle R_{\rm son}\rangle$, time interval between successive plasmoids $\Delta t_{\rm p}$ and asymptotic energy flux $\dot{E}$ (eq.~\ref{Edot}) from constant luminosity models at various values of polar magnetic field strength $B_0$ and rotation rate. These results correspond to a PNS mass of $1.4$\,M$_{\odot}$. Lower mass PNSs  spin down more rapidly, as discussed in Section \ref{MHD Lt}. The quantities have been averaged over time to account for the variation due to plasmoids. $\langle R_{\rm A}\rangle$ and $\langle R_{\rm son}\rangle$ have also been averaged over the latitude $\theta$ since the sonic surfaces are not spherical as can be seen from Figures \ref{map1} and \ref{map2}. For the models in which the sonic points go off the grid near the poles (see Section \ref{MHD results}), the average $\langle R_{\rm A}\rangle$ and $\langle R_{\rm son}\rangle$ include only the magnetosonic points on the grid. In general, $\dot{J}$, $\dot{M}$ and $\dot{E}$ increase with increasing neutrino luminosity, which represents an early phase in the PNS's life after supernova. The spindown time $\tau_{\rm J}$ generally decreases with increasing neutrino luminosity (some exceptions are described in Sections \ref{MHD results} and \ref{MHD Lt}). $\tau_{\rm J}$ is just a few seconds during the early cooling epoch, resulting in rapid spindown of the PNS (see Sections \ref{MHD results} and \ref{MHD Lt} below). 

\subsection{Non-magnetic pure rotation models}
\label{MHD NM}
To test our rotation source term and the boundary conditions, we have run simulations without including the magnetic field for various values of $\Omega_{\star}$. We expect conservation of $\dot{J}$, $\dot{M}$ and $\dot{J}/\dot{M}$ as a function of radius. In the 1D simulations, we find that $\dot{J}/\dot{M}$ is conserved and that it is within $1 \%$ of the expected value $R_{\star}^2 \Omega_{\star}$ for all values of rotation period ($P_{\star}=2\pi/\Omega_{\star}$) ranging from $P_{\star}=1$\,ms to $P_{\star}> 1$\,s. To achieve this level of accuracy, we find that 256 radial zones are sufficient at a period of $1$\,s, while 2048 radial zones are required at a period of $P_{\star}=1$\,ms.  

Conservation of $\dot{J}/\dot{M}$ is observed in 2D non-magnetic simulations as well. For sufficiently slow rotation ($P_{\star}\gtrsim 1$\,s), spherical symmetry is preserved and $\dot{J}/\dot{M}$ is again within $1 \%$ of the expected value $\frac{2}{3}R_{\star}^2\Omega_{\star}$ at a resolution of at least $(N_r,\,N_\theta)=(256,128)$. As expected, spherical symmetry is broken in the models with rapid rotation. For more rapid rotation ($P_{\star}<500$\,ms), $\dot{J}/\dot{M}$ is conserved within 3\% over the radial range up to $500$\,km and within 10\% up to a radius of $1000$\,km at a resolution of at least $(N_r,\,N_\theta)=(512,128)$ for all values of $P_{\star}\geq 50$\,ms. For periods smaller than $50$\,ms, which are not the focus of this paper, conservation of $\dot{J}/\dot{M}$ requires more radial zones (typically 1024 to 2048) using our logarithmic radial zoning and the rotating reference frame.

We have also run non-magnetized pure rotation simulations in the lab frame to compare with the results from the rotating frame. The results agree with each other. The deviation in $\dot{J}/\dot{M}$ between the two frames, at a PNS rotation period of $50$\,ms, is within 5\% up to a radius of $700$\,km and within 10\% up to a radius of $1000$\,km at a resolution of at least $(N_r,\,N_\theta)=(512,128)$. At a given resolution, the results from the rotating frame simulations have systematic deviations from those in the lab frame near the outer boundary. The deviations are due to the large $\phi$-velocity ($= r\Omega_{\star} \sin \theta$) of the grid near the outer boundary. We find that the deviations reduce with increasing radial resolution in the rotating frame simulations. For example, the deviation in $\dot{J}/\dot{M}$ between the two frames at $r\gtrsim 1500$\,km decreases from over 80\% at a resolution of $(N_r,\,N_\theta)=(256,128)$ to under 20\% at a resolution of $(N_r,\,N_\theta)=(512,128)$. These deviations near the outer boundary are why we do not measure integrated quantities too close to the outer boundary (Section \ref{diag}).

\subsection{Magneto-centrifugal models: time-steady snapshots}
\label{MHD results}
Very high magnetic fields ($\sim 10^{15}-10^{16}$\,G) and rapid rotation of the PNS near break-up ($P_{\star}\lesssim 2$\,ms) have the potential to produce GRBs. In this work, we focus on magnetic field strengths $\sim 10^{15}$\,G and slower rotation ($P_{\star}\geq50$\,ms) and study the spindown of the PNS during the cooling epoch. These initial rotation periods are thought to be generic to the normal pulsar population \citep{Faucher2006}. There are a few works that discuss long-term spin evolution and spindown mechanisms of ``normal'' magnetars  (e.g. \citealt{Jawor2022,Malov2022}), and exotic systems like RCW 103 \citep{Ho2017}. Our results contribute towards understanding the spin evolution of relatively slowly rotating magnetars starting from the very early cooling epoch just after birth.

\begin{figure*}
\centering{}
\includegraphics[width=\textwidth]{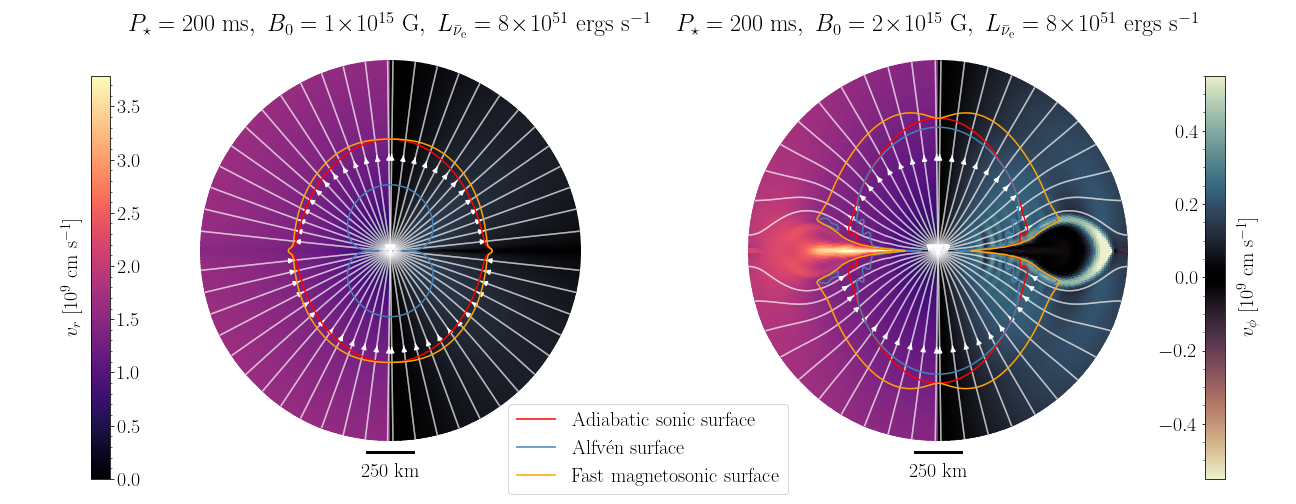}
\caption{2D map of $v_r$ (left half of each panel) and $v_{\phi}$ (right half of each panel) for different values of polar magnetic field at a given rotation period of $200$\,ms and electron type anti-neutrino luminosity of $8\times 10^{51}$\,ergs s$^{-1}$. The left panel corresponds to $B_0=10^{15}$\,G and the right panel corresponds to $B_0=2\times 10^{15}$\,G. We find that plasmoids begin to occur as $B_0$ increases (see Section \ref{MHD results}). The rotation and magnetic axes are along the vertical in the figure. The outer boundary in the figure is at $1000$\,km. The white lines are the magnetic field lines. The method of computation of the magnetosonic surfaces is described in Section \ref{diag}. The slow magnetosonic surface approaches the adiabatic sonic surface for $c_{\rm s} \ll v_{\rm A}$ while it approaches the Alfv\'en surface for $v_{\rm A} \ll c_{\rm s}$.}
\label{map1}
\end{figure*}

Figure \ref{map1} shows a 2D map of $v_r$ and $v_{\phi}$ with the adiabatic sonic surface, the Alfv\'en surface and the fast magnetosonic surface marked. The slow magnetosonic surface approaches the adiabatic sonic surface for $c_{\rm s} \ll v_{\rm A}$ while it approaches the Alfv\'en surface for $v_{\rm A} \ll c_{\rm s}$. The maps correspond to $P_{\star}=200$\,ms and $L_{\rm \bar{\nu}_e}=8 \times 10^{51}$\,ergs s$^{-1}$, which is representative of a cooling PNS on a timescale of $\sim1$\,s. The left panel corresponds to a polar surface magnetic field strength of $B_0=10^{15}$\,G while the right panel corresponds to $B_0=2\times 10^{15}$\,G. The structure of magnetic field outside the PNS is determined by the relative magnitudes of the gas pressure and the polar magnetic field strength $B_0$. Excess gas pressure forces the magnetic field lines to open up into a ``split-monopole'' configuration. A closed zone of the magnetic field forms in regions where the magnetic energy density dominates the gas pressure.

\begin{figure*}
\centering
\includegraphics[width=\textwidth]{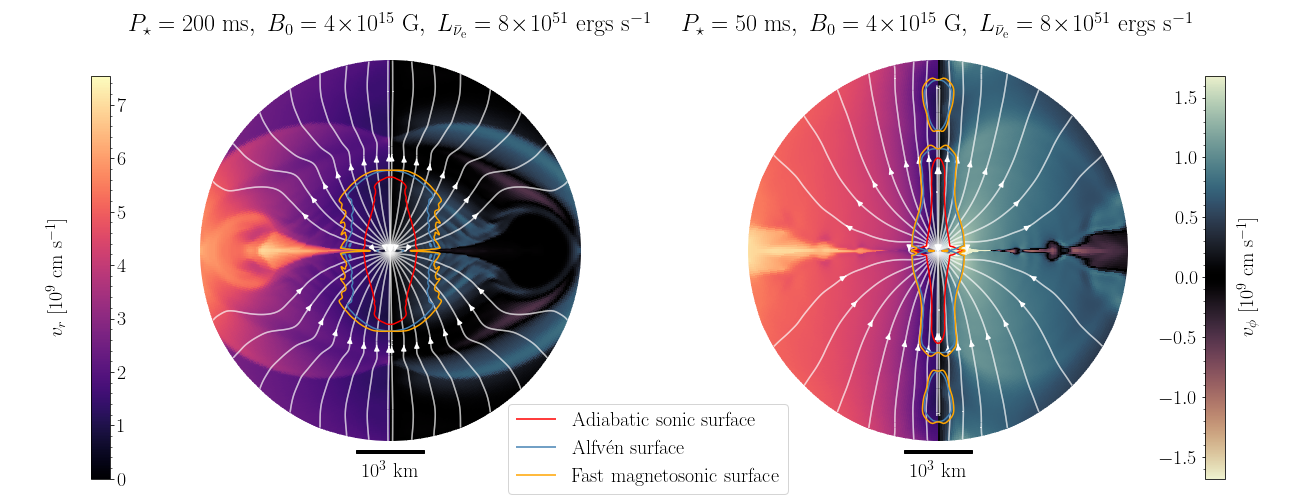}
\caption{2D map of $v_r$ and $v_{\phi}$ similar to Figure \ref{map1} at a polar magnetic field of $B_0=4\times 10^{15}$\,G and electron type anti-neutrino luminosity of $8\times 10^{51}$\,ergs s$^{-1}$. The left panel corresponds to a spin period of $200$\,ms while the right panel corresponds to a spin period of $50$\,ms. As $B_0$ and $\Omega_{\star}$ increase, the sonic surfaces move farther away along the poles. The outer boundary in the figure is at a radius of 2800\,km.} 
\label{map2}
\end{figure*}

\begin{figure*}
\centering
\includegraphics[width=\textwidth]{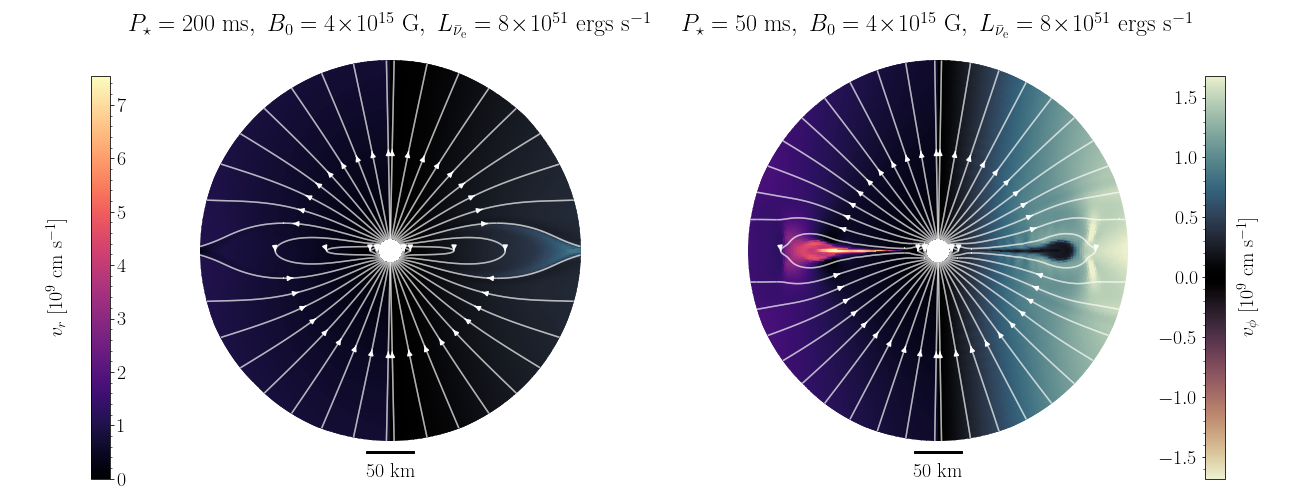}
\caption{Zoomed version of Figure \ref{map2} showing the PNS at the center and the structure of the closed zone of the magnetic field during a plasmoid. The outer boundary in this figure is at a radius of $200$\,km. Here you can see the inner edge of each simulation, i.e. the PNS surface ($R_{\star} = 12$ km).} 
\label{map2_zoom}
\end{figure*}

The left panel in Figure \ref{map1} shows that for $B_0=10^{15}$\,G, the wind reaches a steady state with a large scale split-monopole magnetic field configuration. The structure of models with lower magnetic field strength is qualitatively identical. For a higher magnetic field of $2\times10^{15}$\,G, the configuration does not reach a steady state, but instead exhibits quasi-periodic plasmoid eruptions. The right panel of Figure \ref{map1} shows a snapshot of one such eruption. The closed zone of the magnetic field near the surface of the PNS traps the matter. Eruptions occur when the matter pressure, augmented by neutrino heating exceeds the magnetic tension. After the eruption, the magnetosphere quickly closes again through reconnection. These plasmoid eruptions were predicted on the basis of the neutrino heating and cooling source terms in \cite{Thompson2003} and numerically confirmed in \cite{Thompson2018}. The latter provides a detailed discussion of the physics governing plasmoid eruption, including the predicted timescale. Our magnetized models yield results quantitatively similar to those presented in \cite{Thompson2018}. Plasmoids are potential regions of $r$- or $rp$-process nucleosynthesis because of their very high entropy and rapid expansion timescale (\citealt{Thompson2003,Thompson2018}; see also \citealt{Pruet2006}). We save further details on the properties of the plasmoids and their potential nucleosynthesis for a future paper. Figure \ref{map2} shows 2D maps of $v_r$ and $v_{\phi}$ at a higher magnetic field strength of $B_0=4 \times 10^{15}$\,G. As $B_0$ and $\Omega_{\star}$ increase, the sonic surfaces become more ``cylindrical'' and the sonic points move significantly outwards along the poles, more so during plasmoid eruptions.  Figure \ref{map2_zoom} is a zoomed version of Figure \ref{map2} showing the central PNS and the structure of the closed zone of the magnetic field during a plasmoid and the ``helmet streamer'' type configuration generic to highly-magnetized thermal winds (e.g., \citep{Steinolfson1982,Mestel1987,Endeve2004}).

\begin{figure*}
\centering{}
\includegraphics[width=\textwidth]{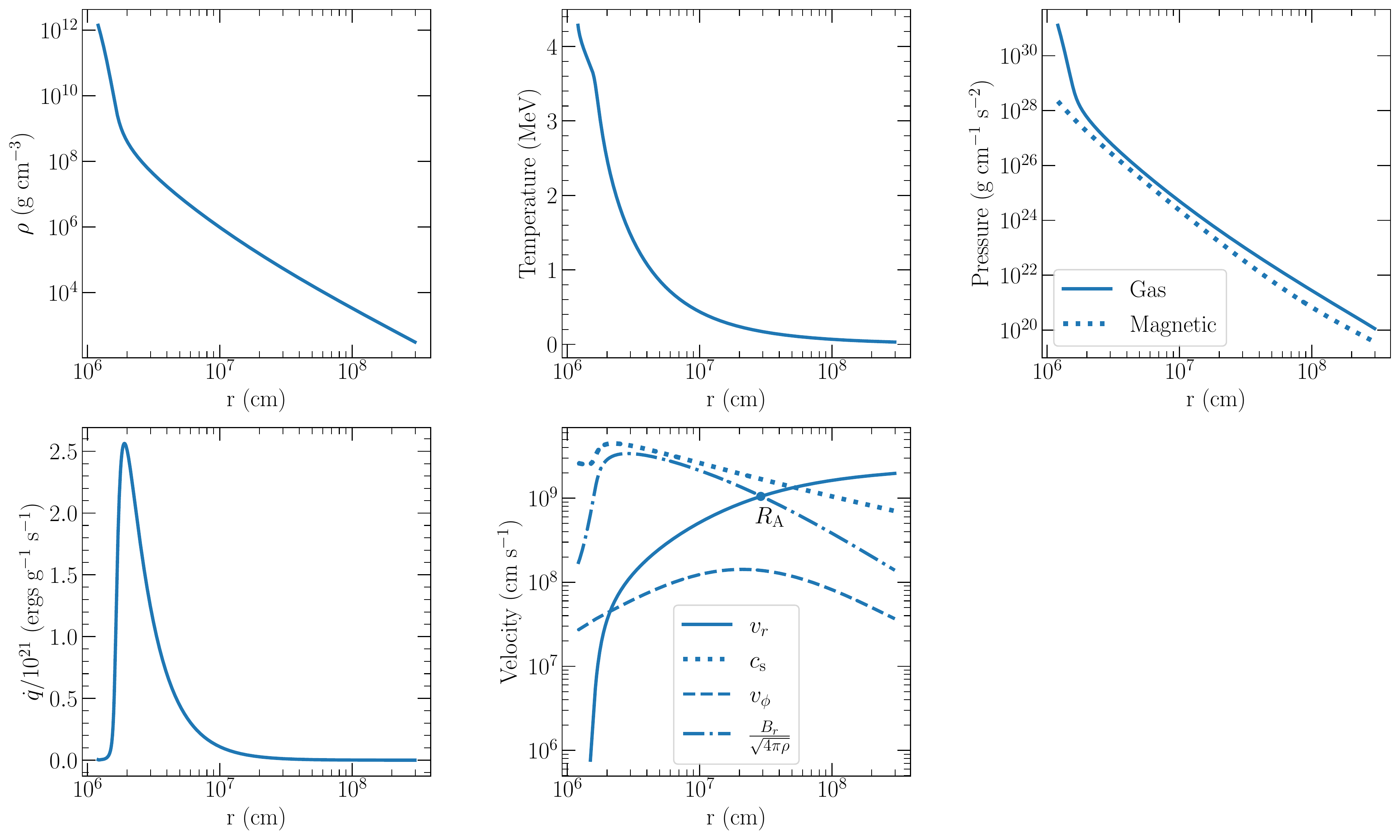}
\caption{1D profiles at $\theta=45 ^{\circ}$ as a function of radius in the simulation with $B_0=10^{15}$\,G, $P_{\star}=200$\,ms and $L_{\rm \bar{\nu}_e}=8\times10^{51}$\,ergs s$^{-1}$. The panels show density, temperature, pressure, specific neutrino heating/cooling rate $\dot{q}$ (related to $\dot{Q}$ in equation \ref{eq:energy} as $\dot{q}=\dot{Q}/ \rho$) and velocity profiles. Since the magnetic pressure is smaller than the gas pressure, plasmoids are not observed with these parameters. The velocity panel also shows the radial Alfv\'en point $R_{\rm A}$, where the radial velocity is equal to the radial Alfv\'en velocity.}
\label{line_plot}
\end{figure*}

Figure \ref{line_plot} shows various physical quantities as a function of radius at $\theta=45^{\circ}$ in the simulation with $P_{\star}=200$\,ms, $B_0=10^{15}$\,G and $L_{\rm \bar{\nu}_e}=8\times10^{51}$\,ergs s$^{-1}$. The corresponding 2D map of the steady-state wind structure is shown in the left panel of Figure \ref{map1}. The bottom second panel in Figure \ref{line_plot} shows the profiles of radial velocity $v_r$, adiabatic sound speed $c_{\rm s}$, radial Alfv\'en velocity and $v_{\phi}$. $v_{\phi}$ increases up to the Alfv\' en radius and then decreases. This shows that the magnetic field forces the wind into effectively co-rotation with the PNS up to the Alfv\'en radius. The top right panel shows the profiles of gas pressure and the magnetic energy density. Since the gas pressure dominates, we expect a steady state with a large scale split-monopole magnetic field configuration with no plasmoids and a very small closed zone within $\sim3$\,km of the PNS surface, as shown in the left panel of Figure \ref{map1}.

\begin{figure*}
\includegraphics[width=\textwidth]{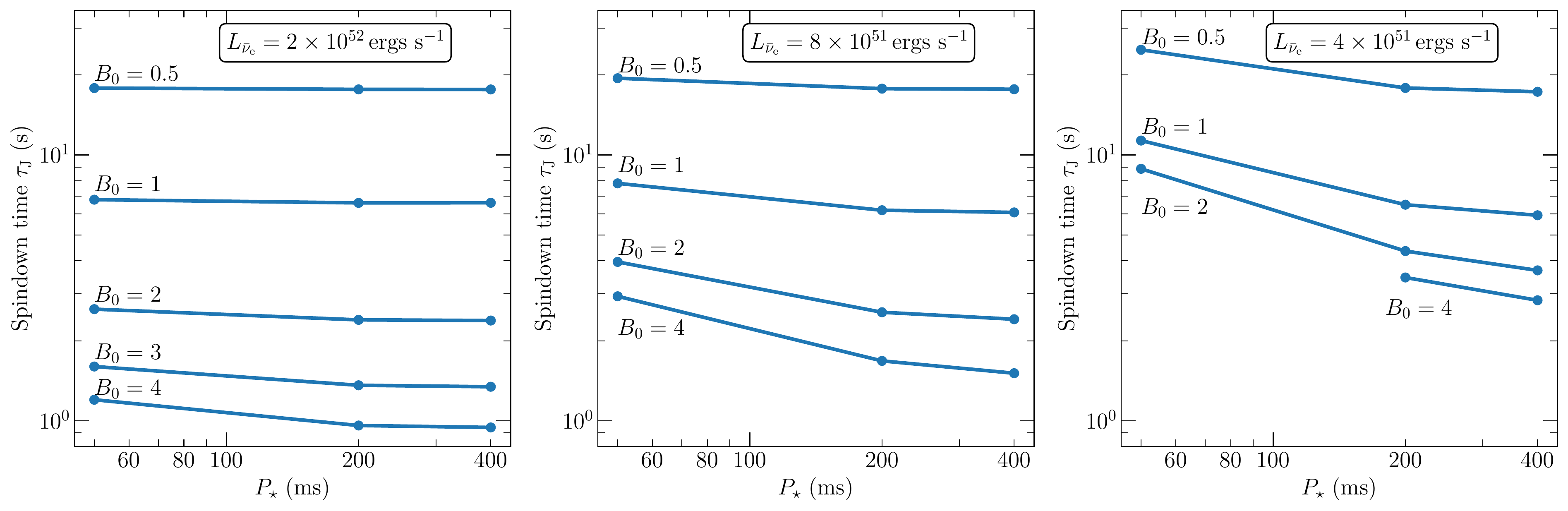}
\caption{Spindown time $\tau_{\rm J}$ as a function of rotation period for different values of $B_0$ (units of $10^{15}$\,G) at (from left to right) $L_{\rm \bar{\nu}_e}=2\times10^{52}$\,ergs s$^{-1}$, $8\times10^{51}$\,ergs s$^{-1}$ and $4\times10^{51}$\,ergs s$^{-1}$. The dots are the actual data points which are connected by continuous lines. Tables \ref{table1}, \ref{table2} and \ref{table3} summarize the values of important physical quantities at these values of $L_{\rm \bar{\nu}_e}$ and $B_0$. The profiles here correspond to a $1.4$\,M$_{\odot}$ PNS. Effect of a different PNS mass is discussed in Section \ref{MHD Lt}.}
\label{tj_Lc}
\end{figure*}

\begin{figure*}
\includegraphics[width=\textwidth]{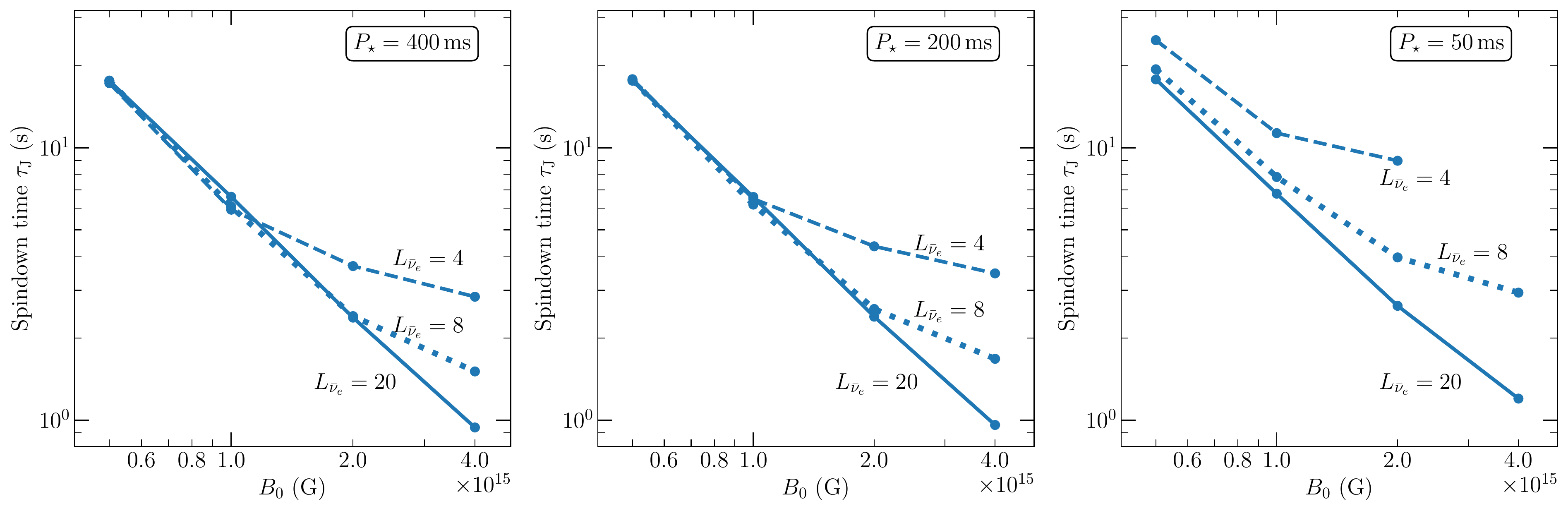}
\caption{Spindown time $\tau_{\rm J}$ as a function of polar magnetic field strength $B_0$  for different values of $L_{\rm \bar{\nu}_e}$ (units of $10^{51}$\,ergs s$^{-1}$) at spin periods of $400$\,ms, $200$\,ms and $50$\,ms. The profiles here correspond to a $1.4$\,M$_{\odot}$ PNS. Effect of a different PNS mass is discussed in Section \ref{MHD Lt}.}
\label{tj_B0}
\end{figure*}

Figures \ref{tj_Lc} and \ref{tj_B0} show $\tau_{\rm J}$ (eq.~\ref{tauj}) as a function of rotation period $P_{\star}$ and polar magnetic field strength $B_0$, respectively, at three different values of $L_{\rm \bar{\nu}_e}$ for a $1.4$\,M$_{\odot}$ PNS at a resolution of $(N_r,\,N_\theta)=(512,256)$. We discuss the effect of a different PNS mass in Section \ref{MHD Lt}. These figures show that for $B_0 \gtrsim 5 \times 10^{14}$\,G, $\tau_{\rm J}$ is less than or of the order the  Kelvin-Helmholtz timescale. During the early cooling period when $L_{\rm \bar{\nu}_e} \gtrsim 10^{52}$\,ergs s$^{-1}$, $\tau_{\rm J}$ is of the order of a few seconds for $B_0 \gtrsim 10^{15}$\,G. We find that the PNS spins down faster as the rotation period increases at a fixed neutrino luminosity and $B_0$, but $\tau_{\rm J}$ is weakly dependent on the spin period for $P_{\star} \gtrsim 200$\,ms at $L_{\rm \bar{\nu}_e} \gtrsim 4 \times 10^{51}$\,ergs s$^{-1}$. From figure \ref{tj_B0}, we find that at relatively low magnetic field strengths of $B_0=5\times 10^{14}$\,G,  $\tau_{\rm J}$ is approximately constant for $P_\star=400$ and 200\,ms as the neutrino luminosity decreases. The near-constant value of $\tau_{\rm J}$ reflects the combination of two physical effects. First, as the neutrino luminosity decreases, the mass-loss rate decreases rapidly because the net neutrino heating at the base of the outflow decreases \citep{QW1996}. Second, as $\dot{M}$ decreases, the value of $R_A$ increases. These two effects work to almost cancel each other so that the angular momentum loss rate $\dot{J}\propto\dot{M}\langle R_A^2\rangle$ of the wind is roughly constant at a given PNS spin period (i.e., $\dot{J}\propto J$). For $P_\star=50$\,ms, the spindown time increases as neutrino luminosity decreases, implying that the decrease in $\dot{M}$ dominates the increase in $R_{\rm A}$. Both the Figures \ref{tj_Lc} and \ref{tj_B0} show that $\tau_{\rm J}$ is strongly dependent on $B_0$, decreasing with increasing $B_0$. This is expected because larger $B_0$ leads to a larger $\langle R_{\rm A}\rangle$. However, as seen in all panels of Figure \ref{tj_B0}, at large values of $B_0$ the decrease in $\tau_{\rm J}$ with $B_0$ becomes less steep. This follows from the fact that as $B_0$ increases, the latitudinal extent of the stable closed zone increases. This region of net zero neutrino heating is held in magnetohydrostatic equilibrium \citep{Thompson2018}, preventing mass loss and lowering $\dot{M}$ overall. Because $\tau_{\rm J}$ is inversely proportional to $\dot{M}$, this effect changes the slope of $\tau_{\rm J}(B_0)$ at fixed neutrino luminosity and spin period. For $P_{\star}=400$ and 200\,ms, Figure \ref{tj_B0} shows that this effect becomes important at $B_0\simeq1$ and $2\times10^{15}$\,G, for $L_{\bar{\nu}_e}=4$ and $8\times10^{51}$\,ergs s$^{-1}$, respectively.

All of the magnetosonic surfaces (described in Section \ref{diag}) should in principal be captured on the computational grid so that the outer boundary conditions have no impact on the calculated wind properties \citep{Metzger2007}. We adjust the radius of the outer boundary $r_{\rm max}$ to capture all the surfaces. The outer boundary has to be moved farther away from the PNS to capture all the magnetosonic surfaces as the neutrino luminosity decreases. This is because the poloidal kinetic energy in the wind decreases with decreasing neutrino luminosity and hence the magnetosonic surfaces move farther away from the PNS. As shown in Figures \ref{map1} and \ref{map2}, as $\Omega_{\star}$ and $B_0$ increase, the sonic surfaces approach a ``cylindrical'' shape and the sonic points along the poles go farther away in radius \citep{Keppens2000}. As a result, the minimum $L_{\rm \bar{\nu}_e}$ achievable in our simulations depends on $P_{\star}$, the polar magnetic field strength $B_0$ and the maximum radius of the grid. We can achieve lower neutrino luminosity for polar magnetic field strengths $B_0 \lesssim 5\times 10^{14}$\,G, but ultimately, we are limited by the magnetosonic points going off the grid. The magnetosonic speeds also  approach the speed of light as the neutrino luminosity decreases, which invalidates our non-relativistic calculations. We will consider further lowering the neutrino luminosity and including relativistic effects in a future work.  

In all the results tabulated here (Tables \ref{table1}-\ref{table4}), the outer boundary is at a radius of $3000$\,km. All the magnetosonic points are on the grid for all values of $B_0$ up to $4\times 10^{15}$\,G considered in this work at $L_{\rm \bar{\nu}_e}=2 \times 10^{52}$\,ergs s$^{-1}$. The sonic points go off the grid only near the poles (mostly during plasmoid eruptions) at $L_{\rm \bar{\nu}_e}=8 \times 10^{51}$\,ergs s$^{-1}$ for $B_0 \geq 4\times10^{15}$\,G and at $L_{\rm \bar{\nu}_e}=4 \times 10^{51}$\,ergs s$^{-1}$ for $B_0\ge2\times10^{15}$\,G. In order to test if the magnetosonic surfaces going off the grid near the poles affects the values of the physical quantities we measure (see Section \ref{diag}), we increased the radius of the outer boundary in some of our calculations to $10000$\,km for $P_{\star}=50$\,ms, 200\,ms and to $15000$\,km for $P_{\star}=400$\,ms so as to capture all the magnetosonic points along the poles. The difference in $\tau_{\rm J}$ is less than $15 \%$ in all the cases with the maximum difference occurring at $L_{\rm \bar{\nu}_e}=4 \times 10^{51}$\,ergs s$^{-1}$ for $B_0=4\times10^{15}$\,G. We note that increasing the radius of the outer boundary does not require the inclusion of relativistic effects because $r_s \Omega_{\star} \sin \theta <c$, where $r_s$ is the radius of a magnetosonic point. We thus stick to a outer boundary radius of 3000\,km to lower the computation time, as increasing the radius of the outer boundary requires a larger number of radial and $\theta$ zones to maintain the required accuracy in the measured physical quantities (see Section \ref{MHD NM}). 

As an aside, we note that approximating the early spindown of the PNS using the standard magnetic dipole formula vastly over-estimates the spindown time. For dipole spindown, the period and the period derivative of a magnetar are related as follows:
\begin{equation}
\label{PPdot}
    \dot{P}_{\star}=kP_{\star}^{2-n},
\end{equation}
where $n$ is the braking index ($=3$ for magnetic dipole spindown) and $k\approx 4\pi^2 B_0^2R_{\star}^6/(6c^3 I)$. We can compare the spindown time predicted by our models and the standard dipole formula for the same parameters. For dipole spindown, the spindown time is given by $\tau_{\rm J}^{\rm d}=P_{\star}/\dot{P}_{\star}=P_{\star}^2/k$. We find that $\tau_{\rm J}^{\rm d}$ is much larger than the spindown time predicted by our models. For example, at $B_0=4\times10^{15}$\,G and $P_{\star}=200$\,ms, $\tau_{\rm J}^{\rm d}=5.6 \times 10^{6}$\,s compared to $\tau_{\rm J}$ of just a few seconds at this magnetic field for all the values neutrino luminosity considered in this paper (see Tables \ref{table1}-\ref{table3}). In contrast to the rapid spindown  predicted by our models (see Section \ref{MHD Lt}), application of the standard dipole formula would predict that the PNS spin period remains constant throughout the cooling epoch.     

\begin{table*}
\centering
    \captionsetup{justification=centering}
	\caption{Wind properties at $L_{\rm \bar{\nu}_e}=2\times10^{52}$\,ergs s$^{-1}$ for a $1.4$\,M$_{\odot}$ PNS. Effect of a different PNS mass is discussed in Section \ref{MHD Lt}}
	\label{table1}
	\begin{tabular}[width=\textwidth]{cccccccccc} 
		\hline
		$L_{\rm \bar{\nu}_e}$ & $B_0$ & $P_{\star}$ & $\tau_{\rm J}$& $\dot{J}$ & $\dot{M}$ &$\langle R_{\rm A} \rangle$ & $\langle R_{\rm son} \rangle$ &$\Delta t_{\rm p}$&$\dot{E}$\\
		($10^{51}$\,ergs s$^{-1}$) & (G) & (ms) & (s) & (g cm$^{2}$\,s$^{-2}$) &(g s$^{-1}$) &(km) & (km) &(ms)&(ergs s$^{-1}$)\\
		\hline\hline \\
		20 & $4 \times 10^{15}$ & 400 &  0.94 &$2.68 \times 10^{46}$  & $3.42 \times 10^{30}$  & 336 & 302 & 25.9 & $2.18 \times 10^{49}$ \\
		& & 200 & 0.96 &$5.26 \times 10^{46}$  & $3.42 \times 10^{30}$  & 339 & 301 & 25.5 & $2.27 \times 10^{49}$ \\
		& & 50 & 1.20 &$1.68 \times 10^{47}$  & $3.46 \times 10^{30}$  & 354 & 282 & 25.4 & $3.76 \times 10^{49}$ \\ \\
		& $3 \times 10^{15}$ & 400 & 1.34 & $1.89 \times 10^{46}$ & $3.55 \times 10^{30}$ & 269 & 297  & 32.5 & $1.90 \times 10^{49}$ \\
		& & 200 & 1.36 & $3.74 \times 10^{46}$ &  $3.55 \times 10^{30}$ & 271 &296 & 32.3 & $1.97 \times 10^{49}$ \\
		& & 50 &  1.60 &$1.26 \times 10^{47}$  & $3.59 \times 10^{30}$  & 280 & 276 & 28.6 & $3.03 \times 10^{49}$ \\ \\
		& $2 \times 10^{15}$ & 400 & 2.38 & $1.06 \times 10^{46}$ & $3.59 \times 10^{30}$ & 197 & 291  & 26.2 & $1.83 \times 10^{49}$ \\
		& & 200 & 2.40 & $2.12 \times 10^{46}$ & $3.60 \times 10^{30}$ & 199 & 291 & 26.1 & $1.87 \times 10^{49}$\\
		& & 50 &  2.63 &$7.69 \times 10^{46}$  & $3.62 \times 10^{30}$  & 202 & 275 & 26.1 & $2.45 \times 10^{49}$ \\ \\
		&$10^{15}$ & 400 & 6.61 & $3.84 \times 10^{45}$ & $3.61 \times 10^{30}$ & 121 & 288 & - & $1.83 \times 10^{49}$ \\
		& & 200 & 6.60 & $7.67 \times 10^{45}$ & $3.61\times10^{30}$ & 121 & 287 &-&$1.83 \times 10^{49}$ \\
		&  & 50 & 6.79 & $2.98 \times 10^{46}$ & $3.63\times10^{30}$ & 120 & 281 &-&$2.02 \times 10^{49}$ \\ \\
		
		& $5 \times 10^{14}$ & 400 & 17.64 & $1.44 \times 10^{45}$ & $3.62 \times 10^{30}$ & 75 & 286 & - &$1.83 \times 10^{49}$  \\
		& & 200 & 17.65 & $2.87 \times 10^{45}$ & $3.62 \times 10^{30}$ & 75 & 286 & - &$1.83 \times 10^{49}$  \\
		& & 50 &  17.85 & $1.14 \times 10^{46}$ & $3.63\times10^{30}$ & 77 & 282 &-&$1.88 \times 10^{49}$ \\ \\
		
		& 0 & 400 &  459.90 & $5.51 \times 10^{43}$ & $3.62\times10^{30}$ & - & 284 &-&$1.83 \times 10^{49}$ \\
		& & 200 & 445.67 & $1.14 \times 10^{44}$ & $3.62\times10^{30}$ & - & 284 &-&$1.83 \times 10^{49}$ \\ 
		& & 50 & 367.36 & $5.52 \times 10^{44}$ & $3.62\times10^{30}$ & - & 284 &-&$1.83 \times 10^{49}$   \\ \\
	\end{tabular}
\end{table*}

\begin{table*}
\centering
    \captionsetup{justification=centering}
	\caption{Wind properties at $L_{\rm \bar{\nu}_e}=8\times10^{51}$\,ergs s$^{-1}$ for a $1.4$\,M$_{\odot}$ PNS. Effect of a different PNS mass is briefly discussed in Section \ref{MHD Lt}.}
	\label{table2}
	\begin{tabular}[width=\textwidth]{cccccccccc} 
		\hline
		$L_{\rm \bar{\nu}_e}$ & $B_0$ & $P_{\star}$ & $\tau_{\rm J}$& $\dot{J}$ & $\dot{M}$ &$\langle R_{\rm A} \rangle$ & $\langle R_{\rm son} \rangle$ &$\Delta t_{\rm p}$&$\dot{E}$\\
		($10^{51}$\,ergs s$^{-1}$) & (G) & (ms) & (s) & (g cm$^{2}$\,s$^{-2}$) &(g s$^{-1}$) &(km) & (km) &(ms)&(ergs s$^{-1}$)\\
		\hline\hline \\
		8 & $4 \times 10^{15}$ & 400 &1.51& $1.68 \times 10^{46}$ & $5.83 \times 10^{29}$ & 836 & 573 & 51.9 & $5.71 \times 10^{48}$\\
		& & 200 & 1.68 & $3.02 \times 10^{46}$ & $5.86 \times 10^{29}$ & 820 & 550 & 50.9 & $6.33 \times 10^{48}$  \\
		& & 50 & 2.94 & $6.95 \times 10^{46}$ & $6.02 \times 10^{29}$ & 736 & 421 & 49.1 & $1.25 \times 10^{49}$ \\ \\

		& $2 \times 10^{15}$ & 400 & 2.41 & $1.05 \times 10^{46}$ & $7.04 \times 10^{29}$ & 484 & 548 & 55.1 & $2.18 \times 10^{48}$  \\
		& & 200& 2.56 & $1.98 \times 10^{46}$ & $7.04 \times 10^{29}$ & 489 & 534 & 54.6 & $2.51 \times 10^{48}$ \\
		& & 50 & 3.96 & $5.12 \times 10^{46}$ & $7.19 \times 10^{29}$ & 501 & 411 & 50.2 & $6.59 \times 10^{48}$ \\ \\
		
		&$10^{15}$ & 400 & 6.08 & $4.17 \times 10^{45}$ & $7.19\times10^{29}$ & 278 & 536 &-&$1.87 \times 10^{48}$  \\
		& & 200 & 6.19 & $8.17 \times 10^{45}$ & $7.20\times10^{29}$ & 290 & 528 &-&$1.99 \times 10^{48}$  \\
		&  & 50 & 7.82 & $2.59 \times 10^{46}$ & $7.31\times10^{29}$ & 284 & 438 &-&$3.69 \times 10^{48}$ \\ \\
		
		& $5 \times 10^{14}$ & 400 & 17.67 & $1.43 \times 10^{45}$ & $7.21\times10^{29}$ & 165 & 531 &-&$1.85 \times 10^{48}$   \\
		& & 200 & 17.76 & $2.85 \times 10^{45}$ & $7.21\times10^{29}$ & 170 & 529 &-&$1.88 \times 10^{48}$   \\
		& & 50 & 19.44 & $1.04 \times 10^{46}$ & $7.28\times10^{29}$ & 171 & 482 &-&$2.42 \times 10^{48}$ \\ \\
		
		& 0 & 400 & 2322.51 & $1.09 \times 10^{43}$ & $7.22\times10^{29}$ & - & 531 &-&$1.84 \times 10^{48}$  \\
		& & 200 & 2309.59 & $2.19 \times 10^{43}$ & $7.22\times10^{29}$ & - & 531 &-&$1.84 \times 10^{48}$  \\ 
		& & 50 & 1941.85 & $1.05 \times 10^{44}$ & $7.22\times10^{29}$ & - & 531 &-&$1.85 \times 10^{48}$   \\ \\
	   
	   \end{tabular}
\end{table*}

\begin{table*}
\centering
    \captionsetup{justification=centering}
	\caption{Wind properties at $L_{\rm \bar{\nu}_e}=4\times10^{51}$\,ergs s$^{-1}$ for a $1.4$\,M$_{\odot}$ PNS. Effect of a different PNS mass is briefly discussed in Section \ref{MHD Lt}.}
	\label{table3}
	\begin{tabular}[width=\textwidth]{cccccccccc} 
		\hline
		$L_{\rm \bar{\nu}_e}$ & $B_0$ & $P_{\star}$ & $\tau_{\rm J}$& $\dot{J}$ & $\dot{M}$ &$\langle R_{\rm A} \rangle$ & $\langle R_{\rm son} \rangle$ &$\Delta t_{\rm p}$&$\dot{E}$\\
		($10^{51}$\,ergs s$^{-1}$) & (G) & (ms) & (s) & (g cm$^{2}$\,s$^{-2}$) &(g s$^{-1}$) &(km) & (km) &(ms)&(ergs s$^{-1}$)\\
		\hline\hline \\
		4 & $4 \times 10^{15}$ & 400  & 2.84 & $9.05\times 10^{45}$ & $1.48\times 10^{29}$ & 1435 & 891 & 140 & $2.0\times 10^{48}$ \\
		& & 200 & 3.46 & $1.49\times 10^{46}$ & $1.51\times 10^{29}$ & 1323 & 772 & 91.6 & $2.37\times 10^{48}$ \\ \\

		& $2 \times 10^{15}$ & 400 & 3.68 & $6.89\times 10^{45}$ & $1.94\times 10^{29}$ & 1043 & 891 & 75.6 & $1.14\times 10^{48}$  \\
		& & 200 & 4.35 & $1.17\times 10^{46}$ & $1.95\times 10^{29}$ & 976 & 787 & 70.7 & $1.33\times 10^{48}$ \\
		& & 50 & 8.87 & $2.30\times 10^{46}$ & $2.01\times 10^{29}$ & 769 & 495 & 68.9 & $3.21\times 10^{48}$ \\ \\
		
		&$10^{15}$ & 400 & 5.93 & $4.27\times 10^{45}$ & $2.11\times 10^{29}$ & 544 & 843 & - & $3.78\times 10^{47}$ \\
		& & 200 & 6.50 & $7.80\times 10^{45}$ & $2.12\times 10^{29}$ & 549 & 774 & - & $4.91\times 10^{47}$  \\
		&  & 50 & 11.33 & $1.79\times 10^{46}$ & $2.18\times 10^{29}$ & 559 & 503 & - & $1.75\times 10^{48}$ \\ \\
		
		& $5 \times 10^{14}$ & 400 & 17.29 & $1.47\times 10^{45}$ & $2.13\times 10^{29}$ & 320 & 849 & - & $3.46\times 10^{47}$  \\
		& & 200 & 17.87 & $2.84\times 10^{45}$ & $2.13\times 10^{29}$ & 319 & 819 & - & $3.79\times 10^{47}$  \\
		& & 50 & 24.87 & $8.15\times 10^{45}$ & $2.17\times 10^{29}$ & 316 & 626 & - & $8.21\times 10^{47}$ \\ \\
		
		& 0 & 400 & 7883.38 & $3.21\times 10^{42}$ & $2.13\times 10^{29}$ & - & 855 & - & $3.34\times 10^{47}$  \\
		& & 200 &  7871.79 & $6.44\times 10^{42}$ & $2.13\times 10^{29}$ & - & 855 & - & $3.34\times 10^{47}$ \\ 
		& & 50  &  7364.81 & $2.76\times 10^{43}$ & $2.13\times 10^{29}$ & - & 854 & - & $3.36\times 10^{47}$  \\ \\
	   
	   \end{tabular}
\end{table*}

\begin{table*}
\centering
    \captionsetup{justification=centering}
	\caption{Wind properties at $L_{\rm \bar{\nu}_e}=2\times10^{51}$\,ergs s$^{-1}$ for a $1.4$\,M$_{\odot}$ PNS. Effect of a different PNS mass is briefly discussed in Section \ref{MHD Lt}.}
	\label{table4}
	\begin{tabular}[width=\textwidth]{cccccccccc} 
		\hline
		$L_{\rm \bar{\nu}_e}$ & $B_0$ & $P_{\star}$ & $\tau_{\rm J}$& $\dot{J}$ & $\dot{M}$ &$\langle R_{\rm A} \rangle$ & $\langle R_{\rm son} \rangle$ &$\Delta t_{\rm p}$&$\dot{E}$\\
		($10^{51}$\,ergs s$^{-1}$) & (G) & (ms) & (s) & (g cm$^{2}$\,s$^{-2}$) &(g s$^{-1}$) &(km) & (km) &(ms)&(ergs s$^{-1}$)\\
		\hline\hline \\
		2 & $5 \times 10^{14}$ & 400  & 18.64 & $1.37\times 10^{45}$ & $6.24\times 10^{28}$ & 510 & 1309 & - & $9.22\times 10^{46}$ \\
		& & 200 & 20.55 & $2.48\times 10^{45}$ & $6.26\times 10^{28}$ & 516 & 1144 & - & $1.21\times 10^{47}$ \\
		& & 50 & 38.99 & $5.20\times 10^{45}$ & $6.45\times 10^{28}$ & 527 & 659 & - & $4.25\times 10^{47}$  \\ \\
		   \end{tabular}
\end{table*}

\begin{figure*}
\centering{}
\includegraphics[width=\textwidth]{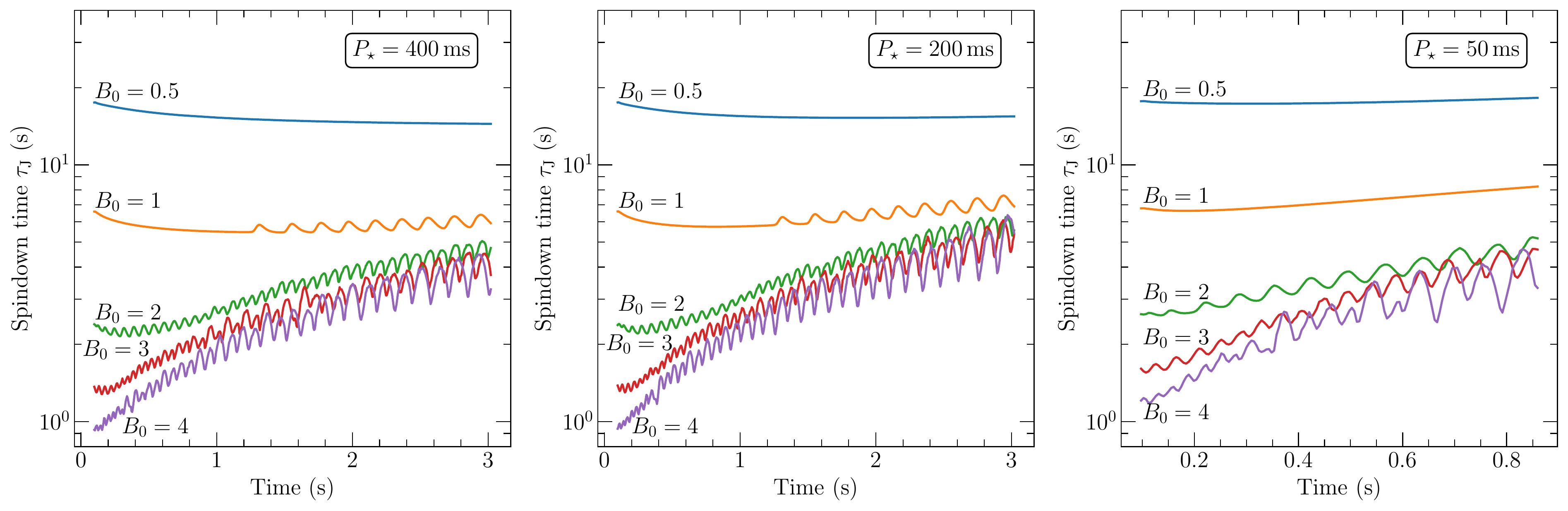}
\caption{Spindown time as a function of time (the start time on the x-axis is the time at which $L_{\rm \bar{\nu}_e}=2 \times 10^{52}$\,ergs s$^{-1}$ in the \citealt{Pons1999} cooling model) for different values of $B_0$ (units of $10^{15}$\,G) at $P_{\star}=400$\,ms, 200\,ms, and 50\,ms (from left to right) for a $1.4$\,M$_{\odot}$ PNS. Effect of a different PNS mass is briefly discussed in Section \ref{MHD Lt}.}
\label{Lt}
\end{figure*}

\subsection{Magneto-centrifugal models: evolution during PNS cooling}
\label{MHD Lt}
The results presented so far give snapshots of the evolution with time-steady boundary conditions. However, for some of the models, the spindown time approaches the instantaneous value of the cooling timescale for the PNS, suggesting that we must be careful in interpreting these results directly.

In order to model the PNS spindown more accurately, we present results from simulations with time dependent neutrino luminosity at the inner boundary to represent the cooling of the PNS during its early stages after birth. We fit the neutrino luminosity data from \cite{Pons1999}. We use $L_{\rm \bar{\nu}_e}(t) \sim t^{-0.56}$ and a constant neutrino mean energy with $\langle\epsilon_{\rm \bar{\nu}_e}\rangle=14$\,MeV and $\langle\epsilon_{\rm \nu_e}\rangle=11$\,MeV as the luminosity evolves during the first $\sim 3$\,s. We start from $L_{\rm \bar{\nu}_e}=2\times10^{52}$\,ergs s$^{-1}$ which corresponds to  $t \sim 0.1$\,s in the \cite{Pons1999} cooling model and we follow the evolution down to a luminosity of $L_{\rm \bar{\nu}_e}=3\times10^{51}$\,ergs s$^{-1}$ at $t=3$\,s, for models with $P_{\star}=200$\,ms and 400\,ms. For $P_{\star}=50$\,ms we evolve down to a luminosity of $L_{\rm \bar{\nu}_e}=6\times10^{51}$\,ergs s$^{-1}$, which corresponds to $t\simeq0.9$\,s in the \cite{Pons1999} cooling models for a 1.4\,M$_\odot$ PNS. The more rapidly rotating models have to be stopped at an earlier time in the evolution because the magnetosonic surfaces cannot be captured on the computational grid as the PNS angular velocity $\Omega_{\star}$ and polar magnetic field $B_0$ increase (see Section \ref{MHD results}). We note that the early luminosity evolution of PNSs is uncertain, and varies between models with different neutrino opacities and different treatments of convection \citep{Pons1999,Roberts2012,Vartanyan2018}. 

Figure \ref{Lt} shows $\tau_{\rm J}$ as a function of time  using the  \cite{Pons1999} cooling model for a $1.4$ M$_{\odot}$ PNS for different values of $B_0$ at \textbf{fixed}  $P_{\star}=400$\,ms, 200\,ms, and 50\,ms, at a resolution of $(N_r,\,N_\theta)=(512,256)$. For this first set of evolutionary models we make the approximation that the PNS spin period does not change during the early evolution. That is, we keep $P_\star$ fixed throughout the few-second evolution in neutrino luminosity even though in principal the PNS is spinning down to longer spin periods. This approximation is justified for the relatively slowly rotating models considered here because $\tau_{\rm J}$ does not depend strongly on $P_\star$. We explicitly test this approximation further below in this sub-section.

At low magnetic field strengths of $B_0=5\times 10^{14}$\,G (top blue lines), Figure \ref{Lt} shows that the spindown time $\tau_{\rm J}$ is approximately constant, but slightly decreasing for $P_\star=400$ and 200\,ms as the PNS cools. This is due to the balance between the increase in $R_{\rm A}$ and the decrease in $\dot{M}$ as the neutrino luminosity decreases, as explained in Section \ref{MHD results}. The slight decrease in $\tau_{\rm J}$ over the 3\,s of evolution for $P_\star=400$ and 200\,ms, reflects the dominance of increase in $R_{\rm A}$. For $P_\star=50$\,ms, the profile is approximately flat during the first second of evolution, implying that the two effects nearly cancel. The time profile of $\tau_{\rm J}$ for all three $P_\star$ values shown is also smooth for $B_0=5\times 10^{14}$\,G, and without the time-dependent modulations seen in the higher magnetic field cases.  

The higher magnetic field cases are qualitatively different from the $B_0=5 \times 10^{14}$\,G and $B_0=10^{15}$\,G cases. First, we see that for each model, $\tau_{\rm J}$ shows modulations that originate from plasmoid eruptions, as shown in Figures \ref{map1}-\ref{map2_zoom}. Spindown is enhanced when the magnetosphere opens during plasmoid eruption.  At low $B_0$ and/or high neutrino luminosity, plasmoids do not develop and the modulations in $\tau_{\rm J}$ are not present. For example, at $B_0=5 \times 10^{14}$\,G, plasmoids do not develop during the evolution shown, but would be expected to emerge as the PNS cools further, beyond what we can currently calculate. In the models with $B_0 = 10^{15}$\,G, plasmoids develop at approximately $1.2$\,s after the start of the evolution, when the neutrino luminosity decreases to the point where the magnetic energy density becomes large compared to the thermal pressure near the base of the outflow. Plasmoids are absent when the gas pressure is always larger than the magnetic energy density (see Section \ref{MHD results}). The time interval between plasmoids at a fixed magnetic field strength and spin period increases as the neutrino luminosity decreases (see Tables \ref{table1}, \ref{table2} and \ref{table3}). For polar magnetic field strength $B_0 \geq 2\times 10^{15}$\,G, we find that the spindown timescale increases as the neutrino luminosity decreases at a fixed value of the spin period and $B_0$. This is in contrast to the $B_0=5\times10^{14}$\,G case where $\tau_{\rm J}$ remains roughly constant during the first $3$\,s of cooling. $\tau_{\rm J}$ increases with decreasing neutrino luminosity for the models with plasmoids because the decrease in $\dot{M}$ dominates the increase in the Alfv\'en radius $R_{\rm A}$. As explained in the previous paragraph, for the models without plasmoids, increase in $R_{\rm A}$ is roughly cancelled by decrease in $\dot{M}$. The larger decrease in $\dot{M}$ with increasing $B_0$ is due to trapping of matter in the closed zone of the magnetic field near the surface of the PNS. Nevertheless, for sufficiently high polar magnetic field $B_0 \gtrsim 2\times{10^{15}}$\,G, the spindown timescale is just a few seconds. 

From the results of Figure \ref{Lt}, we estimate the evolution of the rotation period $P_{\star}(t)$ from $\dot{J}(t)$ as the neutrino luminosity decreases. In the models described in Figure \ref{Lt}, we hold $P_{\star}$ constant as the luminosity evolves in the simulations. As we show below, and as implied by the results of Tables \ref{table1}-\ref{table4}, this is a fairly good approximation since the spindown time $\tau_{\rm J}$ does not sensitively depend on the rotation period of the PNS, at least for $P_{\star}\gtrsim200$\,ms and $L_{\bar{\nu}_{\rm e}}\gtrsim 4 \times 10^{51}$\,ergs s$^{-1}$ (see Figure \ref{tj_Lc}). To account for the spin period dependence of $\dot{J}$, we fit the data obtained from the simulations to get an approximate period scaling: $\dot{J}(t,\alpha) \sim P_{\star}(t)^{-\alpha}$. 
The value of $\alpha$ is obtained by fitting $\dot{J}$ as a function of $P_{\star}$ at different neutrino luminosities and averaging, using the data from Figure \ref{tj_Lc}. From the curve fit and averaging, we get $\alpha\sim 0.75$ for $P_{\star}\gtrsim200$\,ms. For 50\,ms$\lesssim P_{\star} \lesssim200$\,ms, we find $\alpha\sim0.6$. We have, 
\begin{equation}
\label{Jdot_eqn}
    \dot{J}(t,\alpha)=-\frac{2}{5}M_{\star}R_{\star}^2\dot{\Omega}_{\star}(t).
\end{equation}
 where the negative sign accounts for the fact that $\Omega_{\star}$ decreases with time. We assume that $M_{\star}$ and $R_{\star}$ remain constant. A constant $M_{\star}$ is a very good approximation as $\dot{M}$ is very small compared to $M_{\star}$ (see Tables \ref{table1}, \ref{table2}, \ref{table3} and \ref{table4}). The PNS radius decreases by a factor of $\sim 3$ during the first $3$\,s of evolution \citep{Pons1999}. To verify if our calculations of $\tau_{\rm J}$ hold in case of a different PNS radius, we have run a simulation with $L_{\rm \bar{\nu}_e}=2 \times 10^{52}$\,ergs s$^{-1}$, $R_{\star}=20$\,km, $B_0=10^{15}$\,G, $\langle\epsilon_{\rm \bar{\nu}_e}\rangle= 14$\,MeV and $\langle\epsilon_{\rm \nu_e}\rangle=11$\,MeV at a rotation period of $400$\,ms. We find that $\tau_{\rm J}=1.53$\,s at these parameters. $B_0=10^{15}$\,G at $R_{\star}=20$\,km translates to $B_0=2.78\times10^{15}$\,G at $R_{\star}=12$\,km through flux conservation. From Table \ref{table1}, $\tau_{\rm J}=1.34$\,s at $B_0=3\times10^{15}$\,G and $R_{\star}=12$\,km. Hence, we find that the assumption of a constant PNS radius does not give a $\tau_{\rm J}$ which is significantly different compared to a model with evolving PNS radius. An increase in radius of the PNS also increases the moment of inertia, $\dot{J}$ and $\dot{M}$. The effect of a smaller $B_0$ is roughy cancelled by a larger value of $\dot{J}$ at a larger base radius. However, we will try to include a consistent evolution of the PNS radius in a future work.

Separating the spin period dependence and using $\Omega_{\star}(t)=\frac{2\pi}{P_{\star}(t)}$ in equation \ref{Jdot_eqn}, we get,
\begin{equation}
\label{Jdot_p}
    \dot{J}(t)\left(\frac{P_{\star}(t_0)}{P_{\star}(t)}\right)^{\alpha} =\frac{2}{5}M_{\star}R_{\star}^2\frac{2\pi}{P_{\star}^2(t)}\frac{dP_{\star}(t)}{dt}.
\end{equation} 
Integrating the above equation, we get,
\begin{equation}
\label{Pstar_eqn}
    P_{\star}^{\alpha-1}(t)=P_{\star}^{\alpha-1}(t_0)+\frac{5\left(\alpha -1 \right)P_{\star}^{\alpha}(t_0)}{4\pi M_{\star}R_{\star}^2} \int_{t_0}^t \dot{J}(t) dt.
\end{equation}
The integral of $\dot{J}$ is computed by summing the value of $\dot{J}(t)dt$ at snapshots of the simulation separated by 4.8\,ms, which is also the value of $dt$. The timescale for the luminosity to change is $|L_{\rm \bar{\nu}_e}/\dot{L}_{\rm \bar{\nu}_e}| \sim t/0.56$ with our fitted power law to the early time cooling in \cite{Pons1999}. Thus, the outputs have been sampled frequently enough to justify the assumption of constant $\dot{J}$ between successive outputs.

\begin{figure*}
\includegraphics[width=\textwidth]{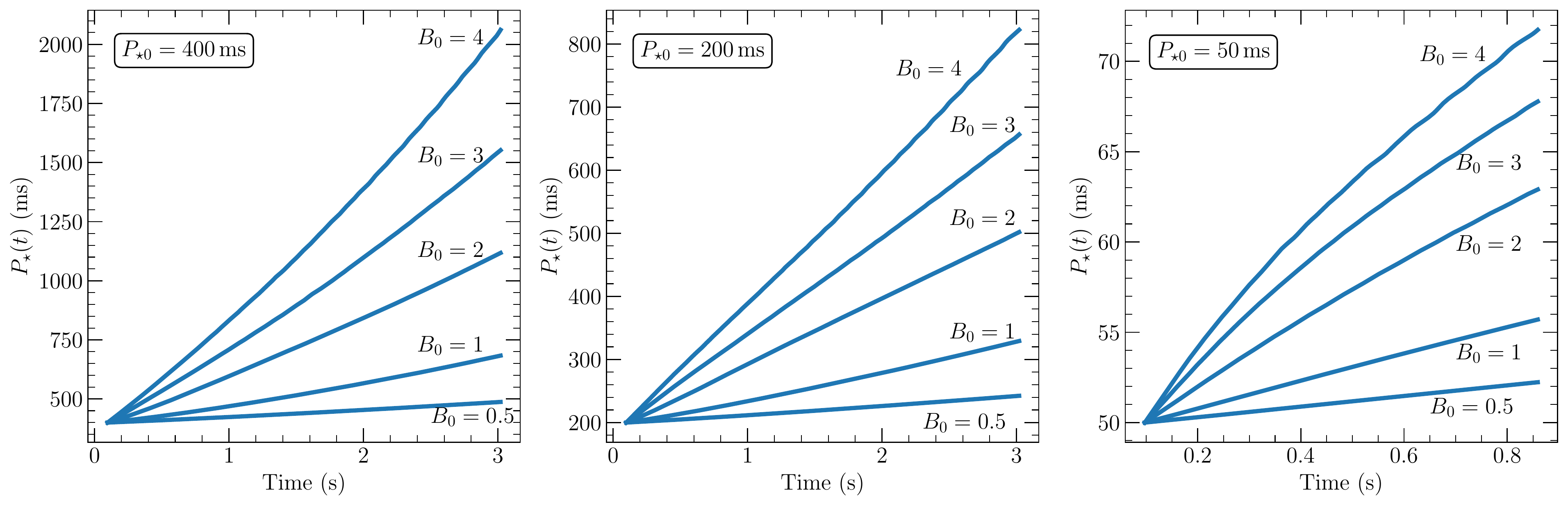}
\caption{Rotation period of the PNS with an initial period of $400$\,ms, $200$\,ms and $50$\,ms (from left to right) as a function of time as the luminosity evolves for different values of $B_0$ (units of $10^{15}$\,G). The neutrino luminosity evolution is obtained from \citealt{Pons1999}. These profiles are from simulations in which we hold the PNS spin period constant at the inner boundary. We use equation \ref{Pstar_eqn} to obtain the spin period as a function of time. We assume a constant PNS radius throughout and a $1.4$\,M$_{\odot}$ PNS. Effect of evolving PNS radius and a different mass is discussed in Section \ref{MHD Lt}.}
\label{Pt}
\end{figure*}

We start from $t_0 \sim 0.1$\,s, which corresponds to the time at which $L_{\rm \bar{\nu}_e}=2\times10^{52}$\,ergs s$^{-1}$ in the \cite{Pons1999} cooling model. $P_{\star}(t_0)$ is the spin period of the PNS at time $t_0$, the initial spin period. Figure \ref{Pt} shows the evolution of $P_{\star}$ with time for $P_{\star}(t_0)=400$, 200 and 50\,ms for the same values of the polar magnetic field strength $B_0$ used in Figure \ref{Lt}. The profiles have been obtained using equation \ref{Pstar_eqn}. We find that the PNS spins down significantly during the first few seconds of the cooling epoch for $B_0\gtrsim2\times10^{15}$\,G for initial spin period $P_{\star}(t_0)\gtrsim200$\,ms. For $P_{\star}(t_0)=50$\,ms, we find that the spindown is not as dramatic, especially for smaller $B_0$, as expected.

\begin{figure*}
\centering{}
\includegraphics[width=\textwidth]{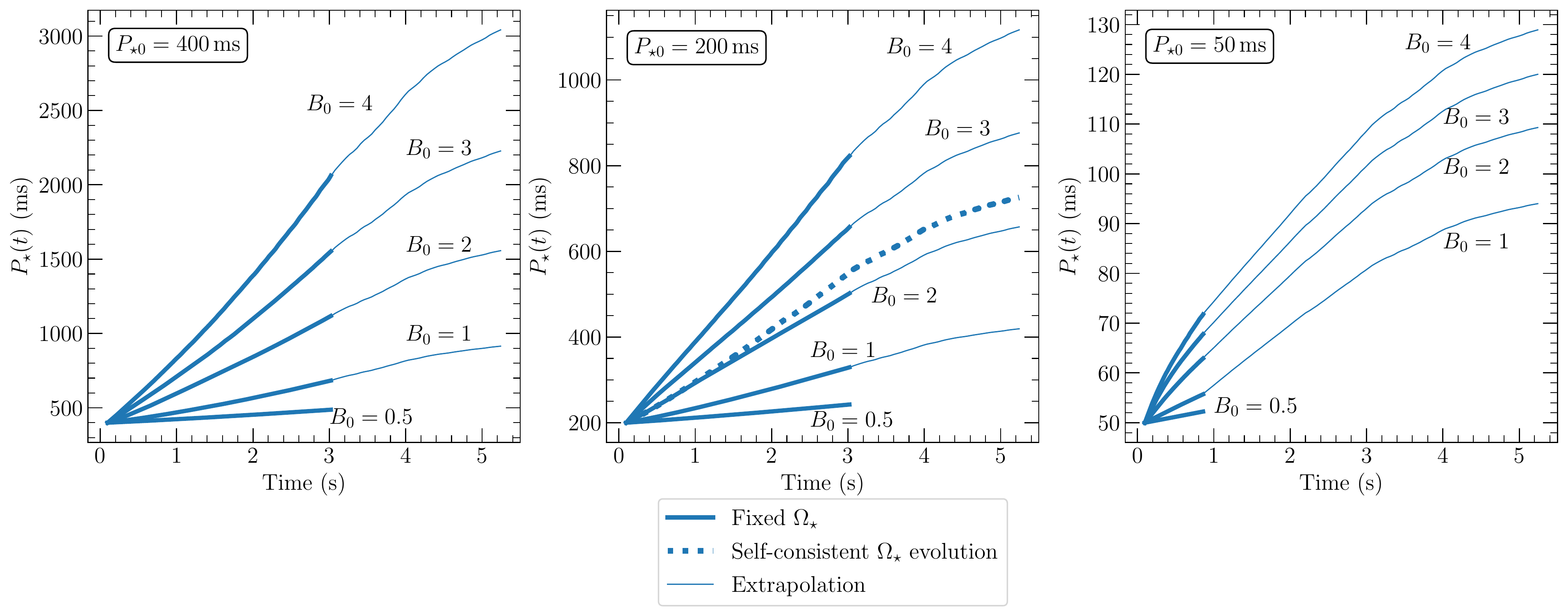}
\caption{This figure is an extension of Figure \ref{Pt}. In this figure, we show the estimates of $P_{\star}(t)$ from the fixed $\Omega_{\star}$ simulations (thick solid lines), self-consistently evolved $P_{\star}(t)$ from the high resolution simulation at $B_0=2\times 10^{15}$\,G (dotted line in the middle panel) and extrapolation of $P_{\star}(t)$ (thin solid lines) assuming $\dot{J}$ is independent of $B_0$ for $B_0\gtrsim10^{15}$\,G at $L_{\rm \bar{\nu}_e}<3\times10^{51}$\,ergs s$^{-1}$ for $P_{\star}>200$\,ms and $L_{\rm \bar{\nu}_e}<6\times10^{51}$\,ergs s$^{-1}$ for 50\,ms $<P_{\star}<150$\,ms. We do not extrapolate the line labelled $B_0=0.5$ because the $\dot{J}$ values for this polar magnetic field strength do not approach the values corresponding to $B_0\geq 10^{15}$\,G (see Figure \ref{Lt}). Evidently, for $B_0=5\times10^{14}$\,G, the spindown is negligible.}
\label{Pt_self}
\end{figure*}

In order to verify the $P_{\star}(t)$ estimates presented in Figure \ref{Pt}, we relax the approximation that the spin period of the PNS is fixed and present a result from a simulation which includes the self-consistent evolution of the PNS spin period $P_{\star}$ with a time-changing inner boundary. This result is from a high resolution simulation with (1024,512) zones with the outer boundary at 10000\,km and polar magnetic field strength $B_0=2 \times10^{15}$\,G. We start from an initial spin period $P_{\star}(t_0)=200$\,ms and $L_{\rm \bar{\nu}_e}=2\times10^{52}$\,ergs s$^{-1}$ . We achieve the self-consistent evolution of the PNS angular velocity by updating the value of $\Omega_{\star}$ at the inner boundary at each timestep using the value of $\dot{J}$ measured at $r= 50$\,km, $\Delta \Omega_{\star}=-\dot{J} dt / I$, where $dt$ is the hydro time-step. We again note that $\dot{J}$ can be measured at any radius not very close to the inner or outer boundaries (see Section \ref{diag}). We assume a constant PNS radius, but in principle, the PNS radius evolves by a factor of $\sim 3$ during the first $\sim 5$\,s cooling epoch \citep{Pons1999}. Since the PNS angular velocity is changing with time, we include the Euler force ($\mathbfit{a}_{\rm Eu}=-\frac{d\boldsymbol{\Omega}_{\star}}{dt} \times \mathbfit{r}$) in the source term and add the corresponding energy to the total energy of the fluid. We follow the evolution of this model up to 5.5\,s corresponding to a luminosity of $L_{\rm \bar{\nu}_e}=1.5\times10^{51}$\,ergs s$^{-1}$ for a $1.4$\,M$_{\odot}$ PNS. We discuss the effect of a different PNS mass in Section \ref{MHD Lt}. As stated in Section \ref{ICs}, the neutrino mean energy is kept constant in time until $3$\,s. For $t>3$\,s, the neutrino mean energy is evolved as a function of time using the \cite{Pons1999} cooling model. The PNS spins down to a period $706$\,ms at the end of $5.5$\,s of evolution. To test the effect of resolution, we have run simulations with the same parameters and self-consistent evolution of the PNS angular velocity at resolutions of $(N_r,\,N_\theta)=(256,128)$ and (256,256). At the end of $5.5$\,s, the PNS spins down to $530$\,ms in the simulation with (256,128) zones, while it spins down to a period of $620$\,ms in the simulation with (256,256) zones. We find that the value of $\dot{J}$ is underestimated in simulations with a low resolution.

In Figure \ref{Pt_self}, we show an extension of Figure \ref{Pt} in which we estimate the evolution of the spin period up to $5.5$\,s based on the high resolution simulation with self-consistent evolution of the spin period. The dotted line in the middle panel in Figure \ref{Pt_self} shows the self-consistent $P_{\star}(t)$ profile at $B_0=2\times10^{15}$\,G. We find that this dotted line is reasonably close to our estimate labelled $B_0=2$ in the middle panel. From Figure \ref{Lt}, we find that at a given spin period, the values of the spindown time $\tau_{\rm J}$ approach each other and hence the value of $\dot{J}$ is roughly independent of $B_0$ for $B_0 \gtrsim 10^{15}$\,G at times $t\gtrsim3$\,s. Using this fact, we extrapolate the spin periods shown in Figure \ref{Pt} for $t>3$\,s, shown by solid thin lines in Figure \ref{Pt_self}, using the values of $\dot{J}$ from the self-consistent simulation at $B_0=2\times 10^{15}$\,G (shown by dotted line in the middle panel in Figure \ref{Pt_self}). We do not extrapolate the line labelled $B_0=0.5$ because the $\dot{J}$ values for this polar magnetic field strength do not approach the values corresponding to $B_0\geq 10^{15}$\,G (see Figure \ref{Lt}). From Figure \ref{Pt} we see that the spindown corresponding to $B_0=5\times10^{14}$\,G is negligible. We assume $\dot{J}$ dependence on the spin period $P_{\star}(t)$ as in equation \ref{Jdot_p} and use $\alpha=0.75$ for $P_{\star}\geq 200$\,ms and $\alpha=0.6$ for $50<P_{\star}<200$\,ms. We use equation \ref{Pstar_eqn} to obtain the values of spin period as a function of time.

We note that the $P_{\star}(t)$ estimates in Figure \ref{Pt_self} for $B_0\gtrsim 2 \times 10^{15}$\,G are likely an underestimate as can be observed from the comparison between the estimate and the actual profile of the spin period at $B_0=2 \times 10^{15}$\,G (middle panel in Figure \ref{Pt_self}). We did not repeat the self-consistent simulations for the other models since these are computationally very expensive, compared to the other simulations.

\begin{figure}
\centering{}
\includegraphics[width=\linewidth]{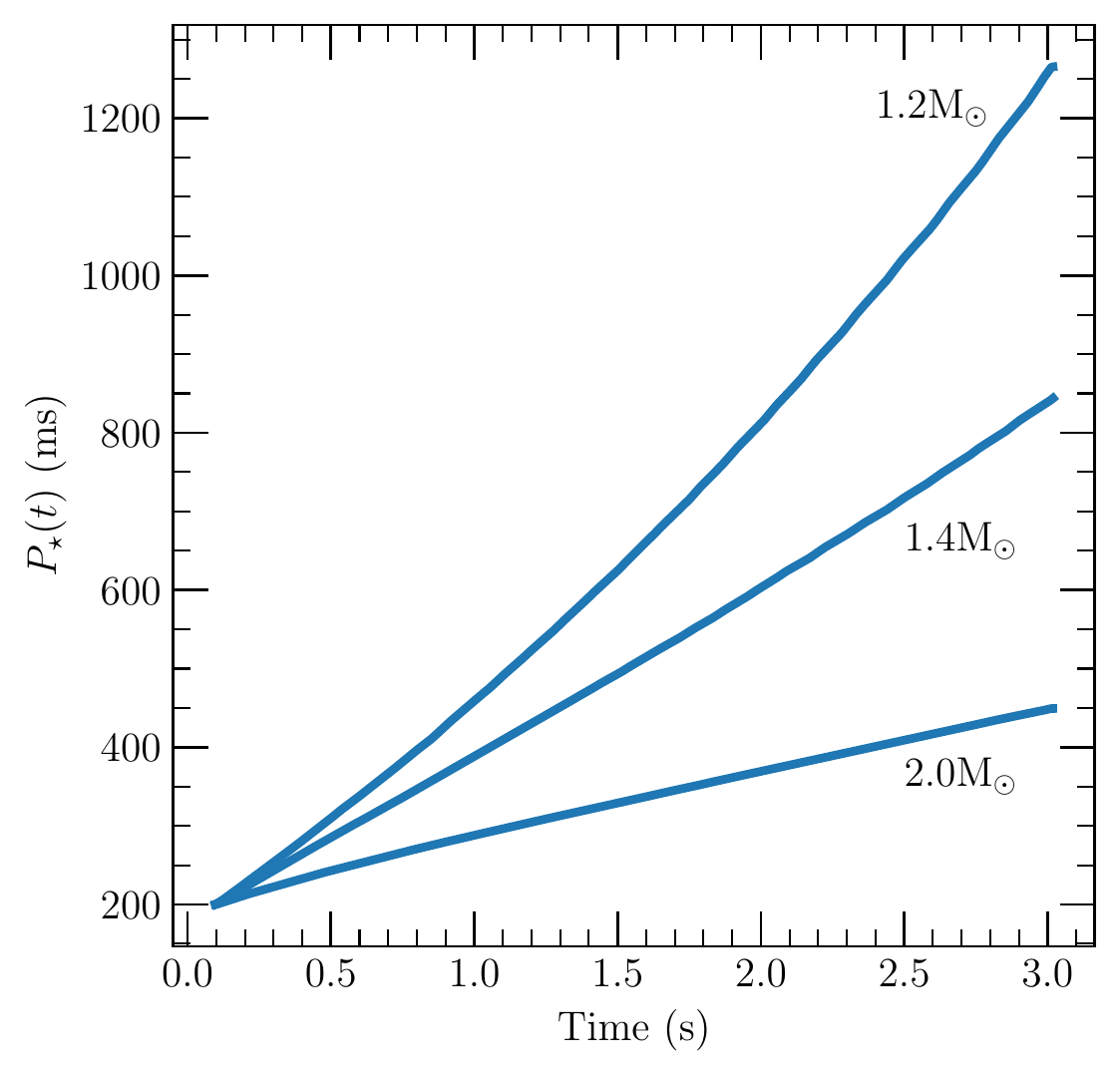}
\caption{Evolution of spin period for various values of PNS mass at $B_0=4\times10^{15}$\,G.}
\label{Pt_mass}
\end{figure}

All the results so far correspond to a PNS of $1.4$ M$_{\odot}$. In order to test the dependence of PNS spindown on PNS mass, we have run simulations with $1.2$ M$_{\odot}$ PNS and $2$ M$_{\odot}$ PNS. The PNS mass affects $\dot{M}$ \citep{QW1996} and the moment of inertia. Figure \ref{Pt_mass} shows the rotation period as a function of time at $B_0=4\times10^{15}$\,G as the PNS spins down for various values of PNS mass. All the profiles are from simulations with self-consistent evolution of the angular velocity $\Omega_{\star}$ of the PNS. We find that the spindown is more efficient for a lower mass PNS. For all the three models, the mean neutrino energy is constant in time and the luminosity profile follows \cite{Pons1999}. For the $1.2$ M$_{\odot}$ model, $\langle\epsilon_{\rm \bar{\nu}_e}\rangle=13.16$\,MeV and $\langle\epsilon_{\rm \nu_e}\rangle=10.34$\,MeV, for the $1.4$ M$_{\odot}$ model, $\langle\epsilon_{\rm \bar{\nu}_e}\rangle=14$\,MeV and $\langle\epsilon_{\rm \nu_e}\rangle=11$\,MeV and for the $2$ M$_{\odot}$ model, $\langle\epsilon_{\rm \bar{\nu}_e}\rangle=15$\,MeV and $\langle\epsilon_{\rm \nu_e}\rangle=11.7$\,MeV. There are two competing effects: the mean neutrino energy and the peak neutrino luminosity decrease with decreasing PNS mass which results in a lower neutrino energy deposition and hence a smaller $\dot{J}$ while the moment of inertia also decreases with decreasing PNS mass. We find that the decrease in the moment of inertia of the PNS with mass wins leading to faster spindown in a PNS with lower mass.

Having the time dependence of the spin period of the PNS, we can construct a metric for ``significant'' spindown during the cooling epoch, at least over the limited early time that we are able to simulate. Defining the critical magnetic field $B_{\rm crit}$ as the value of $B_0$ required for the PNS to spindown to a period of $> 1$\,s at a time $t= 5.5$\,s after the cooling epoch starts, we find from Figure \ref{Pt_self} that $B_{\rm crit} \sim 3.5 \times10^{15}$\,G for $P_{\star0}=200$\,ms and $B_{\rm crit} \sim 1.3 \times10^{15}$\,G for $P_{\star0}=400$\,ms for a $1.4$ M$_{\odot}$ PNS. For a $1.2$ M$_{\odot}$ PNS, we find $B_{\rm crit} \sim 2.5 \times10^{15}$\,G for $P_{\star0}=200$\,ms. Thus, we have the following rough estimate for the $B_{\rm crit}$ scaling:
\begin{equation}
    B_{\rm crit}\sim 1.3\times10^{15}\,{\rm\,G}\left(\frac{P_{\star0}}{400\,{\rm ms}}\right)^{-1.4} \left(\frac{M}{1.4 M_{\odot}}\right)^{2.2}.
    \label{bcrit}
\end{equation}
For $B_0>B_{\rm crit}$, the PNS spins down rapidly during the cooling epoch to spin periods greater than $1$\,s in just $\simeq5.5$\,s of cooling. Because $\tau_{\rm J}$ is less than the cooling time for all $B_0\gtrsim5\times10^{14}$\,G, it is possible that values of $B_0$ less than $B_{\rm crit}$ can lead to spindown of the PNS to periods greater than $1$\,s on longer $\sim10-100$\,s timescales. However, because of the numerical limitations discussed in this sub-section and in Section \ref{section:discussion}, we are currently unable to evolve these models beyond $\sim5-6$\,s.

\section{Discussion \& Conclusions}
\label{section:discussion}

We present a suite of rotating proto-neutron star (PNS) wind calculations to explore the early spin evolution of relatively slowly rotating magnetars in the seconds after birth in core-collapse supernovae. Our simulations span polar magnetic field strengths $B_0$ from $5\times10^{14}$\,G and $4\times10^{15}$\,G, overlapping with the range of inferred magnetic field strengths for Galactic magnetars \citep{Kaspi2017}. We consider initial spin periods in the range inferred for normal Galactic pulsars ($P_{\star}=50-400$\,ms; \citealt{Faucher2006}), consistent with the finding that magnetar-hosting supernova remnants in the Galaxy exhibit normal energetics \citep{Vink2008}, but significantly slower than implied by models of hyper-energetic and super-luminous supernovae and GRBs \citep{Thompson2004,Bucciantini2006,Metzger2007,Kasen2010,Woosely2010}. We follow the PNS evolution over a range of electron type antineutrino luminosity $L_{\rm \bar{\nu}_e}$ between $2\times 10^{52}$\,ergs s$^{-1}$ and $1.5\times 10^{51}$\,ergs s$^{-1}$, which corresponds to a time interval of $\sim 0.1-6$\,s after collapse and explosion \citep{Pons1999}.

Over the range of applicability of our calculations (see below), we find rapid spindown. Using a series of snapshots (see Section \ref{MHD results}), we find that the spindown timescale $\tau_{\rm J}$ (eq.~\ref{tauj}) is less than, or of order, the Kelvin-Helmholz cooling timescale for $B_0\gtrsim5\times10^{14}$\,G, while for $B_0\gtrsim1\times10^{15}$ $\tau_{\rm J}$ is $\sim$\,seconds during the the first few seconds after PNS birth. These calculations indicate that sufficiently magnetic PNSs spin down very rapidly.

Tables \ref{table1}-\ref{table4} list critical diagnostics of the wind, including $\tau_{\rm J}$, $\dot{M}$, $\dot{J}$,  $\dot{E}$, and the average Alfv\'en and adiabatic sonic radii as a function of PNS spin period $P_{\star}$, $B_0$, and neutrino luminosity for a $1.4$\,M$_\odot$ PNS (see Section \ref{diag}).  Figures \ref{tj_Lc} and \ref{tj_B0} summarize these results. For initial spin period $P_{\star}\gtrsim50$\,ms, and $B_0\gtrsim 1-2\times 10^{15}$\,G, we find (see Figure \ref{tj_Lc}) that the  $\tau_{\rm J}\lesssim 3$\,s during the first few seconds of evolution.  These results are generally consistent with previous 1D estimates and calculations \citep{Thompson2004,Metzger2007}. Importantly, despite a factor of $\sim50$ decrease in $\dot{M}$ over the full range of neutrino luminosities covered, because $\langle R_A\rangle^2$ increases as $\dot{M}$ decreases, $\dot{J}$ depends only weakly on $L_{\bar{\nu}_e}$ (see Tables \ref{table1}-\ref{table3}). 

For large enough $B_0$, a ``closed zone" and ``helmet streamer'' configuration develops, which decreases $\dot{M}$ from the PNS. As predicted by \cite{Thompson2003}, 
we find that once a closed zone forms, the outer part is unstable to periodic plasmoid ejections (see Figure \ref{map2_zoom}), 
which modulate $\dot{J}$ and $\dot{E}$. The plasmoid material exhibits high entropy, with implications for heavy-element nucleosynthesis in PNS winds \citep{Thompson2018}.  The thermodynamics and dynamics of the plasmoids have not yet been explored as a function of PNS rotation rate and will be the subject of a future work.

Since the spindown time is of order the cooling time, we construct a series of representative evolutionary models that follow the PNS evolution through the first few seconds. PNS spindown is generic, but the specific details depend on the cooling model. The results we report here correspond to \cite{Pons1999} cooling model, which controls $L_{\bar{\nu}_e}(t)$. Figure \ref{Lt} shows the spindown time $\tau_{\rm J}$ as a function of time at a fixed PNS spin period. Adopting a different PNS cooling model from \cite{Li2021} does not change these results qualitatively, but does change the detailed time evolution of the PNS spin period.  

Figure \ref{Pt_self} shows the evolution of the PNS spin period for different values of $B_0$ as the neutrino luminosity decreases with time (see Section \ref{MHD Lt}). We find that for  $B_0\gtrsim 1.3 \times10^{15}\,{\rm\,G}(P_{\star0}/{400\,\rm\,ms})^{-1.4}(M/1.4{\rm M}_\odot)^{2.2}$ (eq.~\ref{bcrit}) the PNS spins down to a period greater than $1$\,s during the first $\simeq 5-6$\,s of evolution. Even lower values of the magnetic field will likely lead to rapid spindown during the cooling epoch. For example, Figure \ref{Lt} shows that for $B_0=10^{15}$\,G and $5\times10^{14}$\,G, the spindown time can approach the cooling time at $\sim6-10$\,s and $\sim15-20$\,s, respectively, depending on $P_{\star}$. However, our calculations cannot yet explore these late times because of the numerical limitations discussed in Section \ref{MHD Lt} and below. Even so, there exists the possibility that magnetars born with fields of  $B_0\sim5\times10^{14}$\,G can spin down to periods of order seconds during the $\sim10-100$\,s cooling epoch.

Our results may have direct implications for the interpretation of the ages and spindown histories of magnetars in the Galaxy. Based on their measured spin periods and period-derivatives, the  ``characteristic'' age $t^{\prime}_{\rm c}$ of magnetars (or pulsars) is obtained by integrating equation \ref{PPdot} for $P_{\star}$:
\begin{equation}
    \label{act_tc}
    t^{\prime}_{\rm c}= \frac{P_{\star}}{\left(n-1\right)\dot{P}_{\star}}\left[1-\left(\frac{P^{\prime}_{\star}}{P_{\star}}\right)^{n-1}\right],
\end{equation}
where $P^{\prime}_{\star}$ is the spin period of the magnetar at the time dipole radiation begins. Assuming $P^{\prime}_{\star}  \ll  P_{\star}$ and $n=3$, we get the well-known expression for the characteristic age, $t_{\rm c}=P_{\star}/2\dot{P}_{\star}$. However, the assumption $P^{\prime}_{\star}  \ll  P_{\star}$ is not valid for magnetars that undergo rapid spindown during the cooling epoch, before the onset of dipole spindown. In this case, $P_{\star}/2\dot{P}_{\star}$ overestimates the age of the magnetar. Rapid spindown of proto-magnetars during the neutrino cooling epoch may thus help explain the observation that some magnetars have  characteristic ages $t_{\rm c}$ greater than their SN remnant (SNR) ages \citep{Olausen2014}. For example, consider magnetar 1E 1841--045 with $P\simeq11.78$\,s, $B_0\simeq6.9\times10^{14}$\,G, a characteristic age $P_{\star}/2\dot{P}_{\star}=4.7$\,kyr, and an associated SNR age of $0.5-1$\,kyr (see Tables 2 and 7 in \citealt{Olausen2014}).This implies that the factor $\left[1-\left(\frac{P^{\prime}_{\star}}{P_{\star}}\right)^{n-1}\right]$ in equation \ref{act_tc} is $\simeq0.1-0.2$. Assuming $n=3$, the spin period of the magnetar at the onset of dipole spindown is $P^{\prime}_{\star}=10.5-11.2$\,s. One interpretation is that the magnetar was born in the core collapse process with a $\sim10-11$\,s spin period. However, with $B_0\simeq6.9\times10^{14}$\,G, the spindown timescale at birth would be $\lesssim10$\,s for $M\simeq1.4$\,M$_\odot$ (see Figure \ref{Lt}). Thus, given our results, another interpretation is that the magnetar was born with a spin period typical of normal pulsars, but spun down rapidly during the first $\sim10-100$\,s of its existence, before the onset of the dipole spindown phase, to reach a spin period that is very nearly that currently observed. Indeed, the discrepancy between characteristic age $t_{\rm c}$ and the SNR age may provide direct evidence for an early phase of strong spindown.

As a qualitatively different and more problematic example, consider 1E 2259+586, which has $P_\star\simeq6.98$\,s,  $P_{\star}/2\dot{P}_{\star}=230$\,kyr, and SNR age of $14$\,kyr. Assuming $n=3$,  $P^{\prime}_{\star}=6.76$\,s. This might again point to strong early spindown, but the inferred dipole field of 1E 2259+586 is $\simeq6\times10^{13}$\,G, which is sufficiently weak that it should not affect spindown in early phases. However, this argument assumes that the  magnetic field strength is constant over the life of the magnetar. In fact, a number of works instead suggest that the magneto-thermal evolution of magnetars on kyr timescales may lead to lower overall magnetic field strength as magnetars age \citep{Dall'Osso2012,Gao2012}), further complicating the interpretation of magnetar spin evolution. Finally, it is worth noting that there are other magnetars with SNR age greater than their characteristic age \citep{Olausen2014}, which early proto-magnetar spindown cannot explain if the magnetic field strength is constant in time. 

The extreme magnetar E 161348-5055 in the supernova remnant RCW 103 \citep{Luca2006} with a 6.67 hour X-ray periodicity presents a potentially dramatic example of perhaps near-complete spindown. Models of E 161348-5055 (e.g. \citealt{Dong2007}, \citealt{Ho2017}) suggest that its exceptionally long rotation period could have been caused by a fallback accretion disk. Instead, we speculate that the early-time magneto-centrifugal winds we calculate here might be able to nearly stop magnetar rotation during the cooling epoch, as suggested by \cite{Thompson2004}. Our results imply that for $B_0\gtrsim2\times10^{15}$\,G,  $M\lesssim1.4$\,M$_\odot$, and initial spin period $\gtrsim 400$\,ms, rapid spindown during the cooling phase may lead to exceptionally long spin periods. Even though we are forced to extrapolate our results into regions that we cannot yet simulate, Figures \ref{Lt} and \ref{Pt_mass} suggest that $\dot{J}/J\sim{\rm constant}$ and that $\gtrsim10$\,e-foldings of the spin period may be possible during the cooling epoch. If so, this would bring the magnetar to hour- or day-scale spin periods, like E 161348-5055, and perhaps even more speculatively, like the 16 day periodicity seen in the repeating FRB 180916.J0158+65 \citep{Amiri2020}. Perhaps some magnetars undergo complete spindown during the cooling epoch. 

A combination of practical numerical issues currently limits our ability to compute the wind evolution throughout the full $\sim10-100$\,s cooling epoch. As the neutrino luminosity decreases, the gradients at the PNS surface become steep, necessitating high spatial resolution, while the critical magnetosonic surfaces become cylindrical and extend to large scales, moving off the computational domain along the rotation axis and compromising our measurements of the angular momentum and energy loss rates. Additionally, the Alfv\'en velocity approaches $c$ as the PNS cools and the flow becomes Poynting flux-dominated, necessitating a relativistic calculation \citep{Thompson2004, Bucciantini2006,Metzger2007,Metzger2011}. Indeed, the transition from a non-relativistic neutrino-heated magnetocentrifugally driven outflow to a relativistic Poynting flux dominated wind is a hallmark of the early evolution of magnetized neutron stars. To test the assertions and speculations above, it will be necessary to simulate much deeper into the cooling epoch.

In addition to relativistic effects, there are simplifying assumptions in the physics of our calculations that should be noted. We do not follow the PNS evolution from times before the SN explosion, and hence the PNS cooling model is set by hand. The cooling model controls the neutrino luminosities and energies as a function of time, which will at minimum be a function of mass \citep{Pons1999}, and (at rapid rotation rates) the spin period \citep{Thompson2007}. As discussed in Section \ref{section:model}, we hold the electron fraction $Y_{\rm e}$ constant as a function of radius in our simulations, despite the fact that $Y_{\rm e}$ does evolve near the PNS surface before reaching an asymptotic value within a few PNS radii \citep{QW1996}. We further consider a simplified heating/cooling function and  equation of state. In particular, the EOS is accurate only at high temperature $T\gtrsim0.5$\,MeV where electron/positron pairs are relativistic \citep{QW1996}, and does not include $\alpha$ particles. 

In addition to these microphysical pieces, there are macrophysical investigations that are pressing avenues for further work. In particular, we wish to extend our results to the rapidly-spinning magnetars with $P_{\star}\sim1-10$\,ms and $B_0 \gtrsim 1\times10^{14}$\,G that may power SLSNe and long-duration GRBs \citep{Thompson2004,Komissarov2007,Kasen2010, Margalit2018}, and possibly short-duration GRBs formed after NS-NS mergers that produce magnetars \citep{Metzger2008_2,Bucciantini2012,Metzger2018}. Further, for the early times explored, there is the important question of how the wind interacts with the overlying and expanding supernova shockwave itself as it disassembles the inner massive star's core \citep[see e.g.][]{Bucciantini2009}. Finally, in addition to plasmoid eruptions highlighted here  (Section \ref{MHD results}), fallback accretion onto the PNS can open the magnetosphere \citep{Parfrey2016,Metzger2018_2} and maintain the neutrino luminosity \citep{Metzger2018_3}, both of which can lead to faster PNS spindown.

\section*{Acknowledgments}
\label{section:acknowledgements}
We thank Yan-Fei Jiang, Zhaohuan Zhu, Jim Stone, Kengo Tomida, and Adam Finley for helpful discussions. TAT thanks Asif ud-Doula, Brian Metzger, Phil Chang, Niccol\'o Bucciantini, and Eliot Quataert for discussions and collaboration on this and related topics. TP, TAT, and MJR are supported in part by NASA grant 80NSSC20K0531. MC acknowledges support from the U.~S.\ National Science Foundation (NSF) under Grants AST-1714267 and PHY-1804048 (the latter via the Max-Planck/Princeton Center (MPPC) for Plasma Physics). Parts of the results in this work make use of the colormaps in the CMasher package \citep{CMasher}.

\section*{Data Availability}
The implementation of the EOS and the problem generator file to run the simulations using Athena++ are available upon request.


\bibliographystyle{mnras}
\bibliography{ref} 

\begin{thebibliography}{}
\makeatletter
\relax
\def\mn@urlcharsother{\let\do\@makeother \do\$\do\&\do\#\do\^\do\_\do\%\do\~}
\def\mn@doi{\begingroup\mn@urlcharsother \@ifnextchar [ {\mn@doi@}
  {\mn@doi@[]}}
\def\mn@doi@[#1]#2{\def\@tempa{#1}\ifx\@tempa\@empty \href
  {http://dx.doi.org/#2} {doi:#2}\else \href {http://dx.doi.org/#2} {#1}\fi
  \endgroup}
\def\mn@eprint#1#2{\mn@eprint@#1:#2::\@nil}
\def\mn@eprint@arXiv#1{\href {http://arxiv.org/abs/#1} {{\tt arXiv:#1}}}
\def\mn@eprint@dblp#1{\href {http://dblp.uni-trier.de/rec/bibtex/#1.xml}
  {dblp:#1}}
\def\mn@eprint@#1:#2:#3:#4\@nil{\def\@tempa {#1}\def\@tempb {#2}\def\@tempc
  {#3}\ifx \@tempc \@empty \let \@tempc \@tempb \let \@tempb \@tempa \fi \ifx
  \@tempb \@empty \def\@tempb {arXiv}\fi \@ifundefined
  {mn@eprint@\@tempb}{\@tempb:\@tempc}{\expandafter \expandafter \csname
  mn@eprint@\@tempb\endcsname \expandafter{\@tempc}}}

\bibitem[\protect\citeauthoryear{{Barr{\`e}re}, {Guilet}, {Reboul-Salze},
  {Raynaud}  \& {Janka}}{{Barr{\`e}re} et~al.}{2022}]{Barrere2022}
{Barr{\`e}re} P.,  {Guilet} J.,  {Reboul-Salze} A.,  {Raynaud} R.,   {Janka}
  H.~T.,  2022, arXiv e-prints, \href
  {https://ui.adsabs.harvard.edu/abs/2022arXiv220601269B} {p. arXiv:2206.01269}

\bibitem[\protect\citeauthoryear{{Beniamini}, {Hotokezaka}, {van der Horst}  \&
  {Kouveliotou}}{{Beniamini} et~al.}{2019}]{Beniamini2019}
{Beniamini} P.,  {Hotokezaka} K.,  {van der Horst} A.,   {Kouveliotou} C.,
  2019, \mn@doi [\mnras] {10.1093/mnras/stz1391}, \href
  {https://ui.adsabs.harvard.edu/abs/2019MNRAS.487.1426B} {487, 1426}

\bibitem[\protect\citeauthoryear{{Bucciantini}, {Thompson}, {Arons}, {Quataert}
   \& {Del Zanna}}{{Bucciantini} et~al.}{2006}]{Bucciantini2006}
{Bucciantini} N.,  {Thompson} T.~A.,  {Arons} J.,  {Quataert} E.,   {Del Zanna}
  L.,  2006, \mn@doi [\mnras] {10.1111/j.1365-2966.2006.10217.x}, \href
  {http://adsabs.harvard.edu/abs/2006MNRAS.368.1717B} {368, 1717}

\bibitem[\protect\citeauthoryear{{Bucciantini}, {Quataert}, {Arons}, {Metzger}
  \& {Thompson}}{{Bucciantini} et~al.}{2008}]{Bucciantini2008}
{Bucciantini} N.,  {Quataert} E.,  {Arons} J.,  {Metzger} B.~D.,   {Thompson}
  T.~A.,  2008, \mn@doi [\mnras] {10.1111/j.1745-3933.2007.00403.x}, \href
  {http://adsabs.harvard.edu/abs/2008MNRAS.383L..25B} {383, L25}

\bibitem[\protect\citeauthoryear{{Bucciantini}, {Quataert}, {Metzger},
  {Thompson}, {Arons}  \& {Del Zanna}}{{Bucciantini}
  et~al.}{2009}]{Bucciantini2009}
{Bucciantini} N.,  {Quataert} E.,  {Metzger} B.~D.,  {Thompson} T.~A.,  {Arons}
  J.,   {Del Zanna} L.,  2009, \mn@doi [\mnras]
  {10.1111/j.1365-2966.2009.14940.x}, \href
  {http://adsabs.harvard.edu/abs/2009MNRAS.396.2038B} {396, 2038}

\bibitem[\protect\citeauthoryear{{Bucciantini}, {Metzger}, {Thompson}  \&
  {Quataert}}{{Bucciantini} et~al.}{2012}]{Bucciantini2012}
{Bucciantini} N.,  {Metzger} B.~D.,  {Thompson} T.~A.,   {Quataert} E.,  2012,
  \mn@doi [\mnras] {10.1111/j.1365-2966.2011.19810.x}, \href
  {https://ui.adsabs.harvard.edu/abs/2012MNRAS.419.1537B} {419, 1537}

\bibitem[\protect\citeauthoryear{{Burrows} \& {Lattimer}}{{Burrows} \&
  {Lattimer}}{1986}]{Burrows1986}
{Burrows} A.,  {Lattimer} J.~M.,  1986, \mn@doi [\apj] {10.1086/164405}, \href
  {https://ui.adsabs.harvard.edu/abs/1986ApJ...307..178B} {307, 178}

\bibitem[\protect\citeauthoryear{{Burrows}, {Hayes}  \& {Fryxell}}{{Burrows}
  et~al.}{1995}]{Burrows1995}
{Burrows} A.,  {Hayes} J.,   {Fryxell} B.~A.,  1995, \mn@doi [\apj]
  {10.1086/176188}, \href
  {https://ui.adsabs.harvard.edu/abs/1995ApJ...450..830B} {450, 830}

\bibitem[\protect\citeauthoryear{{Cardall} \& {Fuller}}{{Cardall} \&
  {Fuller}}{1997}]{Cardall1997}
{Cardall} C.~Y.,  {Fuller} G.~M.,  1997, \mn@doi [\apjl] {10.1086/310838},
  \href {https://ui.adsabs.harvard.edu/abs/1997ApJ...486L.111C} {486, L111}

\bibitem[\protect\citeauthoryear{{Chime/Frb Collaboration} et~al.,}{{Chime/Frb
  Collaboration} et~al.}{2020}]{Amiri2020}
{Chime/Frb Collaboration} et~al., 2020, \mn@doi [\nat]
  {10.1038/s41586-020-2398-2}, \href
  {https://ui.adsabs.harvard.edu/abs/2020Natur.582..351C} {582, 351}

\bibitem[\protect\citeauthoryear{{Coleman}}{{Coleman}}{2020}]{Coleman2020}
{Coleman} M. S.~B.,  2020, \mn@doi [\apjs] {10.3847/1538-4365/ab82ff}, \href
  {https://ui.adsabs.harvard.edu/abs/2020ApJS..248....7C} {248, 7}

\bibitem[\protect\citeauthoryear{{Dall'Osso}, {Granot}  \& {Piran}}{{Dall'Osso}
  et~al.}{2012}]{Dall'Osso2012}
{Dall'Osso} S.,  {Granot} J.,   {Piran} T.,  2012, \mn@doi [\mnras]
  {10.1111/j.1365-2966.2012.20612.x}, \href
  {https://ui.adsabs.harvard.edu/abs/2012MNRAS.422.2878D} {422, 2878}

\bibitem[\protect\citeauthoryear{{De Luca}, {Caraveo}, {Mereghetti}, {Tiengo}
  \& {Bignami}}{{De Luca} et~al.}{2006}]{Luca2006}
{De Luca} A.,  {Caraveo} P.~A.,  {Mereghetti} S.,  {Tiengo} A.,   {Bignami}
  G.~F.,  2006, \mn@doi [Science] {10.1126/science.1129185}, \href
  {https://ui.adsabs.harvard.edu/abs/2006Sci...313..814D} {313, 814}

\bibitem[\protect\citeauthoryear{{Duncan} \& {Thompson}}{{Duncan} \&
  {Thompson}}{1992}]{Duncan1992}
{Duncan} R.~C.,  {Thompson} C.,  1992, \mn@doi [\apjl] {10.1086/186413}, \href
  {https://ui.adsabs.harvard.edu/abs/1992ApJ...392L...9D} {392, L9}

\bibitem[\protect\citeauthoryear{{Endeve}, {Holzer}  \& {Leer}}{{Endeve}
  et~al.}{2004}]{Endeve2004}
{Endeve} E.,  {Holzer} T.~E.,   {Leer} E.,  2004, \mn@doi [\apj]
  {10.1086/381239}, \href
  {https://ui.adsabs.harvard.edu/abs/2004ApJ...603..307E} {603, 307}

\bibitem[\protect\citeauthoryear{{Faucher-Gigu{\`e}re} \&
  {Kaspi}}{{Faucher-Gigu{\`e}re} \& {Kaspi}}{2006}]{Faucher2006}
{Faucher-Gigu{\`e}re} C.-A.,  {Kaspi} V.~M.,  2006, \mn@doi [\apj]
  {10.1086/501516}, \href
  {https://ui.adsabs.harvard.edu/abs/2006ApJ...643..332F} {643, 332}

\bibitem[\protect\citeauthoryear{{Gao}, {Peng}, {Wang}  \& {Yuan}}{{Gao}
  et~al.}{2012}]{Gao2012}
{Gao} Z.~F.,  {Peng} Q.~H.,  {Wang} N.,   {Yuan} J.~P.,  2012, \mn@doi [\apss]
  {10.1007/s10509-012-1139-x}, \href
  {https://ui.adsabs.harvard.edu/abs/2012Ap&SS.342...55G} {342, 55}

\bibitem[\protect\citeauthoryear{{Ho} \& {Andersson}}{{Ho} \&
  {Andersson}}{2017}]{Ho2017}
{Ho} W. C.~G.,  {Andersson} N.,  2017, \mn@doi [\mnras]
  {10.1093/mnrasl/slw186}, \href
  {https://ui.adsabs.harvard.edu/abs/2017MNRAS.464L..65H} {464, L65}

\bibitem[\protect\citeauthoryear{{Janka} \& {Mueller}}{{Janka} \&
  {Mueller}}{1996}]{Janka1996}
{Janka} H.~T.,  {Mueller} E.,  1996, \aap, \href
  {https://ui.adsabs.harvard.edu/abs/1996A&A...306..167J} {306, 167}

\bibitem[\protect\citeauthoryear{{Jawor} \& {Tauris}}{{Jawor} \&
  {Tauris}}{2022}]{Jawor2022}
{Jawor} J.~A.,  {Tauris} T.~M.,  2022, \mn@doi [\mnras]
  {10.1093/mnras/stab2677}, \href
  {https://ui.adsabs.harvard.edu/abs/2022MNRAS.509..634J} {509, 634}

\bibitem[\protect\citeauthoryear{{Kasen} \& {Bildsten}}{{Kasen} \&
  {Bildsten}}{2010}]{Kasen2010}
{Kasen} D.,  {Bildsten} L.,  2010, \mn@doi [\apj]
  {10.1088/0004-637X/717/1/245}, \href
  {https://ui.adsabs.harvard.edu/abs/2010ApJ...717..245K} {717, 245}

\bibitem[\protect\citeauthoryear{{Kaspi} \& {Beloborodov}}{{Kaspi} \&
  {Beloborodov}}{2017}]{Kaspi2017}
{Kaspi} V.~M.,  {Beloborodov} A.~M.,  2017, \mn@doi [\araa]
  {10.1146/annurev-astro-081915-023329}, \href
  {https://ui.adsabs.harvard.edu/abs/2017ARA&A..55..261K} {55, 261}

\bibitem[\protect\citeauthoryear{{Keppens} \& {Goedbloed}}{{Keppens} \&
  {Goedbloed}}{2000}]{Keppens2000}
{Keppens} R.,  {Goedbloed} J.~P.,  2000, \mn@doi [\apj] {10.1086/308395}, \href
  {https://ui.adsabs.harvard.edu/abs/2000ApJ...530.1036K} {530, 1036}

\bibitem[\protect\citeauthoryear{{Komissarov} \& {Barkov}}{{Komissarov} \&
  {Barkov}}{2007}]{Komissarov2007}
{Komissarov} S.~S.,  {Barkov} M.~V.,  2007, \mn@doi [\mnras]
  {10.1111/j.1365-2966.2007.12485.x}, \href
  {https://ui.adsabs.harvard.edu/abs/2007MNRAS.382.1029K} {382, 1029}

\bibitem[\protect\citeauthoryear{{Li}}{{Li}}{2007}]{Dong2007}
{Li} X.-D.,  2007, \mn@doi [\apjl] {10.1086/521791}, \href
  {https://ui.adsabs.harvard.edu/abs/2007ApJ...666L..81L} {666, L81}

\bibitem[\protect\citeauthoryear{{Li}, {Roberts}  \& {Beacom}}{{Li}
  et~al.}{2021}]{Li2021}
{Li} S.~W.,  {Roberts} L.~F.,   {Beacom} J.~F.,  2021, \mn@doi [\prd]
  {10.1103/PhysRevD.103.023016}, \href
  {https://ui.adsabs.harvard.edu/abs/2021PhRvD.103b3016L} {103, 023016}

\bibitem[\protect\citeauthoryear{{Malov} \& {Marozava}}{{Malov} \&
  {Marozava}}{2022}]{Malov2022}
{Malov} I.~F.,  {Marozava} H.~P.,  2022, \mn@doi [Astronomy Reports]
  {10.1134/S106377292202007X}, \href
  {https://ui.adsabs.harvard.edu/abs/2022ARep...66...25M} {66, 25}

\bibitem[\protect\citeauthoryear{{Margalit}, {Metzger}, {Berger}, {Nicholl},
  {Eftekhari}  \& {Margutti}}{{Margalit} et~al.}{2018}]{Margalit2018}
{Margalit} B.,  {Metzger} B.~D.,  {Berger} E.,  {Nicholl} M.,  {Eftekhari} T.,
   {Margutti} R.,  2018, \mn@doi [\mnras] {10.1093/mnras/sty2417}, \href
  {https://ui.adsabs.harvard.edu/abs/2018MNRAS.481.2407M} {481, 2407}

\bibitem[\protect\citeauthoryear{{Mereghetti}, {Pons}  \&
  {Melatos}}{{Mereghetti} et~al.}{2015}]{Mereghetti2015}
{Mereghetti} S.,  {Pons} J.~A.,   {Melatos} A.,  2015, \mn@doi [\ssr]
  {10.1007/s11214-015-0146-y}, \href
  {https://ui.adsabs.harvard.edu/abs/2015SSRv..191..315M} {191, 315}

\bibitem[\protect\citeauthoryear{{Mestel} \& {Spruit}}{{Mestel} \&
  {Spruit}}{1987}]{Mestel1987}
{Mestel} L.,  {Spruit} H.~C.,  1987, \mn@doi [\mnras] {10.1093/mnras/226.1.57},
  \href {https://ui.adsabs.harvard.edu/abs/1987MNRAS.226...57M} {226, 57}

\bibitem[\protect\citeauthoryear{{Metzger}, {Thompson}  \&
  {Quataert}}{{Metzger} et~al.}{2007}]{Metzger2007}
{Metzger} B.~D.,  {Thompson} T.~A.,   {Quataert} E.,  2007, \mn@doi [\apj]
  {10.1086/512059}, \href
  {https://ui.adsabs.harvard.edu/abs/2007ApJ...659..561M} {659, 561}

\bibitem[\protect\citeauthoryear{{Metzger}, {Quataert}  \&
  {Thompson}}{{Metzger} et~al.}{2008a}]{Metzger2008_2}
{Metzger} B.~D.,  {Quataert} E.,   {Thompson} T.~A.,  2008a, \mn@doi [\mnras]
  {10.1111/j.1365-2966.2008.12923.x}, \href
  {https://ui.adsabs.harvard.edu/abs/2008MNRAS.385.1455M} {385, 1455}

\bibitem[\protect\citeauthoryear{{Metzger}, {Thompson}  \&
  {Quataert}}{{Metzger} et~al.}{2008b}]{Metzger2008}
{Metzger} B.~D.,  {Thompson} T.~A.,   {Quataert} E.,  2008b, \mn@doi [\apj]
  {10.1086/526418}, \href {http://adsabs.harvard.edu/abs/2008ApJ...676.1130M}
  {676, 1130}

\bibitem[\protect\citeauthoryear{{Metzger}, {Giannios}, {Thompson},
  {Bucciantini}  \& {Quataert}}{{Metzger} et~al.}{2011}]{Metzger2011}
{Metzger} B.~D.,  {Giannios} D.,  {Thompson} T.~A.,  {Bucciantini} N.,
  {Quataert} E.,  2011, \mn@doi [\mnras] {10.1111/j.1365-2966.2011.18280.x},
  \href {https://ui.adsabs.harvard.edu/abs/2011MNRAS.413.2031M} {413, 2031}

\bibitem[\protect\citeauthoryear{{Metzger}, {Thompson}  \&
  {Quataert}}{{Metzger} et~al.}{2018a}]{Metzger2018}
{Metzger} B.~D.,  {Thompson} T.~A.,   {Quataert} E.,  2018a, \mn@doi [\apj]
  {10.3847/1538-4357/aab095}, \href
  {https://ui.adsabs.harvard.edu/abs/2018ApJ...856..101M} {856, 101}

\bibitem[\protect\citeauthoryear{{Metzger}, {Beniamini}  \&
  {Giannios}}{{Metzger} et~al.}{2018b}]{Metzger2018_2}
{Metzger} B.~D.,  {Beniamini} P.,   {Giannios} D.,  2018b, \mn@doi [\apj]
  {10.3847/1538-4357/aab70c}, \href
  {https://ui.adsabs.harvard.edu/abs/2018ApJ...857...95M} {857, 95}

\bibitem[\protect\citeauthoryear{{Metzger}, {Beniamini}  \&
  {Giannios}}{{Metzger} et~al.}{2018c}]{Metzger2018_3}
{Metzger} B.~D.,  {Beniamini} P.,   {Giannios} D.,  2018c, \mn@doi [\apj]
  {10.3847/1538-4357/aab70c}, \href
  {https://ui.adsabs.harvard.edu/abs/2018ApJ...857...95M} {857, 95}

\bibitem[\protect\citeauthoryear{{Olausen} \& {Kaspi}}{{Olausen} \&
  {Kaspi}}{2014}]{Olausen2014}
{Olausen} S.~A.,  {Kaspi} V.~M.,  2014, \mn@doi [\apjs]
  {10.1088/0067-0049/212/1/6}, \href
  {https://ui.adsabs.harvard.edu/abs/2014ApJS..212....6O} {212, 6}

\bibitem[\protect\citeauthoryear{{Otsuki}, {Tagoshi}, {Kajino}  \&
  {Wanajo}}{{Otsuki} et~al.}{2000}]{Otsuki2000}
{Otsuki} K.,  {Tagoshi} H.,  {Kajino} T.,   {Wanajo} S.-y.,  2000, \mn@doi
  [\apj] {10.1086/308632}, \href
  {https://ui.adsabs.harvard.edu/abs/2000ApJ...533..424O} {533, 424}

\bibitem[\protect\citeauthoryear{{Parfrey}, {Spitkovsky}  \&
  {Beloborodov}}{{Parfrey} et~al.}{2016}]{Parfrey2016}
{Parfrey} K.,  {Spitkovsky} A.,   {Beloborodov} A.~M.,  2016, \mn@doi [\apj]
  {10.3847/0004-637X/822/1/33}, \href
  {https://ui.adsabs.harvard.edu/abs/2016ApJ...822...33P} {822, 33}

\bibitem[\protect\citeauthoryear{{P{\'e}tri}}{{P{\'e}tri}}{2022}]{Petri2022}
{P{\'e}tri} J.,  2022, \mn@doi [\aap] {10.1051/0004-6361/202142522}, \href
  {https://ui.adsabs.harvard.edu/abs/2022A&A...659A.147P} {659, A147}

\bibitem[\protect\citeauthoryear{{Pons} \& {Perna}}{{Pons} \&
  {Perna}}{2011}]{Pons2011}
{Pons} J.~A.,  {Perna} R.,  2011, \mn@doi [\apj] {10.1088/0004-637X/741/2/123},
  \href {https://ui.adsabs.harvard.edu/abs/2011ApJ...741..123P} {741, 123}

\bibitem[\protect\citeauthoryear{{Pons}, {Reddy}, {Prakash}, {Lattimer}  \&
  {Miralles}}{{Pons} et~al.}{1999}]{Pons1999}
{Pons} J.~A.,  {Reddy} S.,  {Prakash} M.,  {Lattimer} J.~M.,   {Miralles}
  J.~A.,  1999, \mn@doi [\apj] {10.1086/306889}, \href
  {https://ui.adsabs.harvard.edu/abs/1999ApJ...513..780P} {513, 780}

\bibitem[\protect\citeauthoryear{{Pruet}, {Hoffman}, {Woosley}, {Janka}  \&
  {Buras}}{{Pruet} et~al.}{2006}]{Pruet2006}
{Pruet} J.,  {Hoffman} R.~D.,  {Woosley} S.~E.,  {Janka} H.~T.,   {Buras} R.,
  2006, \mn@doi [\apj] {10.1086/503891}, \href
  {https://ui.adsabs.harvard.edu/abs/2006ApJ...644.1028P} {644, 1028}

\bibitem[\protect\citeauthoryear{{Qian} \& {Woosley}}{{Qian} \&
  {Woosley}}{1996}]{QW1996}
{Qian} Y.~Z.,  {Woosley} S.~E.,  1996, \mn@doi [\apj] {10.1086/177973}, \href
  {https://ui.adsabs.harvard.edu/abs/1996ApJ...471..331Q} {471, 331}

\bibitem[\protect\citeauthoryear{{Raynaud}, {Guilet}, {Janka}  \&
  {Gastine}}{{Raynaud} et~al.}{2020}]{Raynaud2020}
{Raynaud} R.,  {Guilet} J.,  {Janka} H.-T.,   {Gastine} T.,  2020, \mn@doi
  [Science Advances] {10.1126/sciadv.aay2732}, \href
  {https://ui.adsabs.harvard.edu/abs/2020SciA....6.2732R} {6, eaay2732}

\bibitem[\protect\citeauthoryear{{Roberts}, {Shen}, {Cirigliano}, {Pons},
  {Reddy}  \& {Woosley}}{{Roberts} et~al.}{2012}]{Roberts2012}
{Roberts} L.~F.,  {Shen} G.,  {Cirigliano} V.,  {Pons} J.~A.,  {Reddy} S.,
  {Woosley} S.~E.,  2012, \mn@doi [\prl] {10.1103/PhysRevLett.108.061103},
  \href {https://ui.adsabs.harvard.edu/abs/2012PhRvL.108f1103R} {108, 061103}

\bibitem[\protect\citeauthoryear{{Salmonson} \& {Wilson}}{{Salmonson} \&
  {Wilson}}{1999}]{Salmonson1999}
{Salmonson} J.~D.,  {Wilson} J.~R.,  1999, \mn@doi [\apj] {10.1086/307232},
  \href {https://ui.adsabs.harvard.edu/abs/1999ApJ...517..859S} {517, 859}

\bibitem[\protect\citeauthoryear{{Steinolfson}, {Suess}  \& {Wu}}{{Steinolfson}
  et~al.}{1982}]{Steinolfson1982}
{Steinolfson} R.~S.,  {Suess} S.~T.,   {Wu} S.~T.,  1982, \mn@doi [\apj]
  {10.1086/159872}, \href
  {https://ui.adsabs.harvard.edu/abs/1982ApJ...255..730S} {255, 730}

\bibitem[\protect\citeauthoryear{{Stone}, {Tomida}, {White}  \&
  {Felker}}{{Stone} et~al.}{2019}]{Athena++}
{Stone} J.~M.,  {Tomida} K.,  {White} C.,   {Felker} K.~G.,  2019, {Athena++:
  Radiation GR magnetohydrodynamics code} (\mn@eprint {ascl} {1912.005})

\bibitem[\protect\citeauthoryear{{Sukhbold} \& {Thompson}}{{Sukhbold} \&
  {Thompson}}{2017}]{Sukhbold2017}
{Sukhbold} T.,  {Thompson} T.~A.,  2017, \mn@doi [\mnras]
  {10.1093/mnras/stx2004}, \href
  {https://ui.adsabs.harvard.edu/abs/2017MNRAS.472..224S} {472, 224}

\bibitem[\protect\citeauthoryear{{Thompson}}{{Thompson}}{1994}]{Thompson1994}
{Thompson} C.,  1994, \mn@doi [\mnras] {10.1093/mnras/270.3.480}, \href
  {https://ui.adsabs.harvard.edu/abs/1994MNRAS.270..480T} {270, 480}

\bibitem[\protect\citeauthoryear{{Thompson}}{{Thompson}}{2003}]{Thompson2003}
{Thompson} T.~A.,  2003, \mn@doi [\apjl] {10.1086/374261}, \href
  {http://adsabs.harvard.edu/abs/2003ApJ...585L..33T} {585, L33}

\bibitem[\protect\citeauthoryear{{Thompson}}{{Thompson}}{2007}]{Thompson2007}
{Thompson} T.~A.,  2007, in {Sato} K.,  {Hisano} J.,  eds, Energy Budget in the
  High Energy Universe. pp 251--260 (\mn@eprint {arXiv} {astro-ph/0608231}),
  \mn@doi{10.1142/9789812708342_0028}

\bibitem[\protect\citeauthoryear{{Thompson} \& {Duncan}}{{Thompson} \&
  {Duncan}}{1993}]{Thompson1993}
{Thompson} C.,  {Duncan} R.~C.,  1993, \mn@doi [\apj] {10.1086/172580}, \href
  {https://ui.adsabs.harvard.edu/abs/1993ApJ...408..194T} {408, 194}

\bibitem[\protect\citeauthoryear{{Thompson} \& {ud-Doula}}{{Thompson} \&
  {ud-Doula}}{2018}]{Thompson2018}
{Thompson} T.~A.,  {ud-Doula} A.,  2018, \mn@doi [\mnras]
  {10.1093/mnras/sty480}, \href
  {http://adsabs.harvard.edu/abs/2018MNRAS.476.5502T} {476, 5502}

\bibitem[\protect\citeauthoryear{{Thompson}, {Burrows}  \& {Meyer}}{{Thompson}
  et~al.}{2001}]{Thompson2001}
{Thompson} T.~A.,  {Burrows} A.,   {Meyer} B.~S.,  2001, \mn@doi [\apj]
  {10.1086/323861}, \href
  {https://ui.adsabs.harvard.edu/abs/2001ApJ...562..887T} {562, 887}

\bibitem[\protect\citeauthoryear{{Thompson}, {Chang}  \& {Quataert}}{{Thompson}
  et~al.}{2004}]{Thompson2004}
{Thompson} T.~A.,  {Chang} P.,   {Quataert} E.,  2004, \mn@doi [\apj]
  {10.1086/421969}, \href {http://adsabs.harvard.edu/abs/2004ApJ...611..380T}
  {611, 380}

\bibitem[\protect\citeauthoryear{{Timmes} \& {Swesty}}{{Timmes} \&
  {Swesty}}{2000}]{Timmes2000}
{Timmes} F.~X.,  {Swesty} F.~D.,  2000, \mn@doi [\apjs] {10.1086/313304}, \href
  {https://ui.adsabs.harvard.edu/abs/2000ApJS..126..501T} {126, 501}

\bibitem[\protect\citeauthoryear{{Usov}}{{Usov}}{1992}]{Usov1992}
{Usov} V.~V.,  1992, \mn@doi [\nat] {10.1038/357472a0}, \href
  {https://ui.adsabs.harvard.edu/abs/1992Natur.357..472U} {357, 472}

\bibitem[\protect\citeauthoryear{{Vartanyan}, {Burrows}, {Radice}, {Skinner}
  \& {Dolence}}{{Vartanyan} et~al.}{2018}]{Vartanyan2018}
{Vartanyan} D.,  {Burrows} A.,  {Radice} D.,  {Skinner} M.~A.,   {Dolence} J.,
  2018, \mn@doi [\mnras] {10.1093/mnras/sty809}, \href
  {https://ui.adsabs.harvard.edu/abs/2018MNRAS.477.3091V} {477, 3091}

\bibitem[\protect\citeauthoryear{{Vidotto}, {Jardine}, {Morin}, {Donati},
  {Opher}  \& {Gombosi}}{{Vidotto} et~al.}{2014}]{Vidotto2014}
{Vidotto} A.~A.,  {Jardine} M.,  {Morin} J.,  {Donati} J.~F.,  {Opher} M.,
  {Gombosi} T.~I.,  2014, \mn@doi [\mnras] {10.1093/mnras/stt2265}, \href
  {https://ui.adsabs.harvard.edu/abs/2014MNRAS.438.1162V} {438, 1162}

\bibitem[\protect\citeauthoryear{{Vigan{\`o}}, {Rea}, {Pons}, {Perna},
  {Aguilera}  \& {Miralles}}{{Vigan{\`o}} et~al.}{2013}]{Vigano2013}
{Vigan{\`o}} D.,  {Rea} N.,  {Pons} J.~A.,  {Perna} R.,  {Aguilera} D.~N.,
  {Miralles} J.~A.,  2013, \mn@doi [\mnras] {10.1093/mnras/stt1008}, \href
  {https://ui.adsabs.harvard.edu/abs/2013MNRAS.434..123V} {434, 123}

\bibitem[\protect\citeauthoryear{{Vink}}{{Vink}}{2008}]{Vink2008}
{Vink} J.,  2008, \mn@doi [Advances in Space Research]
  {10.1016/j.asr.2007.06.042}, \href
  {https://ui.adsabs.harvard.edu/abs/2008AdSpR..41..503V} {41, 503}

\bibitem[\protect\citeauthoryear{{Vlasov}, {Metzger}  \& {Thompson}}{{Vlasov}
  et~al.}{2014}]{Vlasov2014}
{Vlasov} A.~D.,  {Metzger} B.~D.,   {Thompson} T.~A.,  2014, \mn@doi [\mnras]
  {10.1093/mnras/stu1667}, \href
  {https://ui.adsabs.harvard.edu/abs/2014MNRAS.444.3537V} {444, 3537}

\bibitem[\protect\citeauthoryear{{Vlasov}, {Metzger}, {Lippuner}, {Roberts}  \&
  {Thompson}}{{Vlasov} et~al.}{2017}]{Vlasov2017}
{Vlasov} A.~D.,  {Metzger} B.~D.,  {Lippuner} J.,  {Roberts} L.~F.,
  {Thompson} T.~A.,  2017, \mn@doi [\mnras] {10.1093/mnras/stx478}, \href
  {https://ui.adsabs.harvard.edu/abs/2017MNRAS.468.1522V} {468, 1522}

\bibitem[\protect\citeauthoryear{{Wanajo}, {Kajino}, {Mathews}  \&
  {Otsuki}}{{Wanajo} et~al.}{2001}]{Wanajo2001}
{Wanajo} S.,  {Kajino} T.,  {Mathews} G.~J.,   {Otsuki} K.,  2001, \mn@doi
  [\apj] {10.1086/321339}, \href
  {https://ui.adsabs.harvard.edu/abs/2001ApJ...554..578W} {554, 578}

\bibitem[\protect\citeauthoryear{{Wheeler}, {Yi}, {H{\"o}flich}  \&
  {Wang}}{{Wheeler} et~al.}{2000}]{Wheeler2000}
{Wheeler} J.~C.,  {Yi} I.,  {H{\"o}flich} P.,   {Wang} L.,  2000, \mn@doi
  [\apj] {10.1086/309055}, \href
  {https://ui.adsabs.harvard.edu/abs/2000ApJ...537..810W} {537, 810}

\bibitem[\protect\citeauthoryear{{White}, {Burrows}, {Coleman}  \&
  {Vartanyan}}{{White} et~al.}{2022}]{White2022}
{White} C.~J.,  {Burrows} A.,  {Coleman} M. S.~B.,   {Vartanyan} D.,  2022,
  \mn@doi [\apj] {10.3847/1538-4357/ac4507}, \href
  {https://ui.adsabs.harvard.edu/abs/2022ApJ...926..111W} {926, 111}

\bibitem[\protect\citeauthoryear{{Woosley}}{{Woosley}}{2010}]{Woosely2010}
{Woosley} S.~E.,  2010, \mn@doi [\apjl] {10.1088/2041-8205/719/2/L204}, \href
  {https://ui.adsabs.harvard.edu/abs/2010ApJ...719L.204W} {719, L204}

\bibitem[\protect\citeauthoryear{{Zhang} \& {M{\'e}sz{\'a}ros}}{{Zhang} \&
  {M{\'e}sz{\'a}ros}}{2001}]{Zhang2001}
{Zhang} B.,  {M{\'e}sz{\'a}ros} P.,  2001, \mn@doi [\apjl] {10.1086/320255},
  \href {https://ui.adsabs.harvard.edu/abs/2001ApJ...552L..35Z} {552, L35}

\bibitem[\protect\citeauthoryear{{Zhu}, {Jiang}, {Baehr}, {Youdin}, {Armitage}
  \& {Martin}}{{Zhu} et~al.}{2021}]{Zhu2021}
{Zhu} Z.,  {Jiang} Y.-F.,  {Baehr} H.,  {Youdin} A.~N.,  {Armitage} P.~J.,
  {Martin} R.~G.,  2021, \mn@doi [\mnras] {10.1093/mnras/stab2517}, \href
  {https://ui.adsabs.harvard.edu/abs/2021MNRAS.508..453Z} {508, 453}

\bibitem[\protect\citeauthoryear{van~der Velden}{van~der
  Velden}{2020}]{CMasher}
van~der Velden E.,  2020, \mn@doi [Journal of Open Source Software]
  {10.21105/joss.02004}, 5, 2004

\makeatother
\end{thebibliography}




\appendix
\section{Inner boundary conditions}\label{app:BCs}
We start from the conservation form of the momentum equation \ref{eq:momentum} and use the following vector identity:
\begin{equation}
\label{app1}
    \frac{1}{\rho}\nabla \cdot \left(\rho \mathbfit{v}\mathbfit{v}\right)=\frac{1}{\rho}\mathbfit{v}\left(\nabla \cdot \left(\rho\mathbfit{v}\right)\right)+\left(\mathbfit{v} \cdot \nabla\right)\mathbfit{v}.
\end{equation}
For time independent $\rho$, using equation \ref{eq:continuity} in equation \ref{app1}, we have,
\begin{equation}
        \frac{1}{\rho}\nabla \cdot \left(\rho \mathbfit{v}\mathbfit{v}\right)= \left(\mathbfit{v} \cdot \nabla\right)\mathbfit{v}.
\end{equation}
Using the same vector identity for $\mathbfit{B}$ as in equation \ref{app1}, we have,
\begin{equation}\label{beq}
    \nabla\left(\frac{B^2}{2}\right) - \nabla \cdot \left(\mathbfit{B}\mathbfit{B}\right)=\nabla\left(\frac{B^2}{2}\right)-\left(\mathbfit{B} \cdot \nabla\right)\mathbfit{B},
\end{equation}
where we have used the fact that divergence of $\mathbfit{B}$ is zero.\\
We have,
\begin{equation}
    -\left(\mathbfit{B} \cdot \nabla\right)\mathbfit{B}=\mathbfit{B}\times \left(\nabla \times \mathbfit{B}\right)-\nabla\left(\frac{B^2}{2}\right),
\end{equation}
where $B=\left|\mathbfit{B}\right|$. The magnetic field is roughly dipolar near the PNS surface. Thus, we have $\mathbfit{B}\times \left(\nabla \times \mathbfit{B}\right)=0$. As a result, the left hand side of equation \ref{beq} reduces to zero. Thus, the time steady momentum equation reduces to
\begin{equation}
\label{grad_eq}
    \frac{1}{\rho}\nabla P=-\frac{GM_{\star}}{r^2}\boldsymbol{\hat{r}}-\left(\mathbfit{v} \cdot \nabla\right)\mathbfit{v}.
\end{equation}
The temperature is nearly constant near the surface of the PNS \citep{QW1996, Thompson2001} and $Y_{\rm e}$ is small near the surface \citep{Thompson2001}. Thus, $\frac{1}{\rho}\nabla P \approx \frac{kT_0}{m_n}\frac{\nabla \rho}{\rho}$ using the equation of state as in \cite{QW1996}, where $T_0$ is the temperature at the base of the PNS set by equating the neutrino heating and cooling rates (see Section \ref{BCs}).

In the rotating reference frame, the $\phi$ velocity of matter at the surface of the PNS is zero. Hence the velocity vector can be written as $\mathbfit{v}^{\prime}=\left(v_r,v_{\theta},0\right)$ in the rotating frame and $\mathbfit{v}=\left(v_r,v_{\theta},r \Omega_{\star} \sin\theta\right )$ in the lab frame. Thus, the $\phi$ derivative of $\rho$ from equation \ref{grad_eq} can be written as
\begin{equation}
   \frac{1}{\rho r \sin \theta}\frac{\partial \rho}{\partial \phi}=-2\Omega_{\star}\left(v_r\sin \theta +v_{\theta}\cos \theta\right)\frac{m_{\rm n}}{kT_0}. 
\end{equation}
The above equation has to vanish to satisfy the axisymmetry condition. So, we set $v_r=0$ and $v_{\theta}=0$ at the inner boundary. With these conditions enforced, the other two partial differential equations from equation \ref{grad_eq} are as follows:
\begin{align}
    \frac{1}{\rho}\frac{\partial \rho}{\partial r}&= \left(r\Omega_{\star}^2\sin^2 \theta -\frac{GM_{\star}}{r^2}\right)\frac{m_{\rm n}}{kT_0}\\
    \frac{1}{\rho}\frac{\partial \rho}{\partial \theta}&=\left(r^2\Omega_{\star}^2\sin \theta \cos \theta \right)\frac{m_{\rm n}}{kT_0}.
\end{align}
The solution to these equations is the density boundary condition \ref{densBC}.\\
\bsp	
\label{lastpage}
\end{document}